\documentclass[12pt,a4paper,english,onecolumn]{IEEEtran}
\usepackage[T1]{fontenc}
\usepackage[latin9]{inputenc}
\usepackage{color}
\usepackage{babel}
\usepackage{verbatim}
\usepackage{float}
\usepackage{mathrsfs}
\usepackage{amsthm}
\usepackage{amsmath}
\usepackage{amssymb}
\usepackage{graphicx}
\usepackage{setspace}
\usepackage{esint}
\usepackage{array}
\usepackage{multirow}
\usepackage{tabularx}
\usepackage{float}
\usepackage{booktabs}
\restylefloat{figure or table}
\usepackage[unicode=true,pdfusetitle,
 bookmarks=false,
 breaklinks=false,pdfborder={0 0 1},backref=false,colorlinks=true]
 {hyperref}
\hypersetup{
 linkcolor=blue, citecolor=red}

\makeatletter
\pdfpageheight\paperheight
\pdfpagewidth\paperwidth

\floatstyle{ruled}
\newfloat{algorithm}{tbp}{loa}
\providecommand{\algorithmname}{Algorithm}
\floatname{algorithm}{\protect\algorithmname}

\theoremstyle{plain}
\newtheorem{thm}{\protect\theoremname}
\theoremstyle{plain}
\newtheorem{lem}[thm]{\protect\lemmaname}

\usepackage{subfigure}
\usepackage{epstopdf}

\makeatother

\providecommand{\lemmaname}{Lemma}
\providecommand{\theoremname}{Theorem}

\begin{document}

\title{Performance Impact of LOS and NLOS Transmissions in Dense Cellular Networks under Rician Fading}

\author{\noindent {\normalsize{}Amir H. Jafari{$^{1}$}, }{\normalsize{} Ming Ding{$^{2}$}, }{\normalsize{} David L$\acute{\textrm{o}}$pez-P$\acute{\textrm{e}}$rez{$^{3}$},
}{\normalsize{} Jie Zhang{$^{1}$}}
\\
{$~^{1}$}{\normalsize{}Department of Electronic $\&$ Electrical Engineering, University of Sheffield, UK}\\
{$~^{2}$}{\normalsize{}Data61, CSIRO, Australia}\\
{$~^{3}$}{\normalsize{}Nokia Bell Labs, Ireland} 
}
\maketitle
\begin{abstract}

In this paper, we analyse the performance of dense small cell network (SCNs). 
We derive analytical expressions for both their coverage probability and their area spectral efficiency (ASE) 
using a path loss model that considers both line-of-sight (LOS) and non-LOS (NLOS) components. 
Due to the close proximity of small cell base stations (BSs) and user equipments (UEs) in such dense SCNs, 
we also consider Rician fading as the multi-path fading channel model for both the LOS and NLOS fading transmissions.
The Rayleigh fading used in most of existing works analysing dense SCNs is not accurate enough. 
Then, we compare the performance impact of LOS and NLOS transmissions in dense SCNs under Rician fading with that based on Rayleigh fading. 
The analysis and the simulation results show that in dense SCNs where LOS transmissions dominate the performance, 
the impact of Rician fading on the overall system performance is minor,
and does not help to address the performance losses brought by the transition of many interfering signals from NLOS to LOS. 

\end{abstract}


\section{Introduction\label{sec:Introduction}}

Dense small cell networks (SCNs) hold the promise to rapidly increase the network capacity for the fifth generation (5G) of cellular communications~\cite{Report_CISCO} 
by deploying base stations (BSs) much closer to user equipments (UEs) and reusing the spectrum intensively~\cite{Tutor_smallcell}.
However, the small distances between transmitters and receivers also bring a change in the channel characteristics, 
which in turn may significantly impact the network performance.
For example, the channel may become line-of-sight (LOS) dominated with the smaller distances between BSs and UEs,
with the subsequent loss of channel diversity. 
Unfortunately, most of the previous theoretical studies on dense SCNs have neglected these facts, 
and do not capture these important channel properties.
Some works for instance do not consider the probability of line-of-sight (LOS).
Instead, they use path loss models that do not differentiate between LOS and Non-LOS (NLOS) transmissions,
and other simplifications,
which may have led to misleading conclusions, 
or conclusions that do not apply to the full spectrum of BS densities,
e.g., the coverage probability is independent of the number of deployed small cell BSs in interference-limited fully-loaded cellular networks,
and/or the area spectral efficiency (ASE) will linearly increase with network densification~\cite{JSAC_Dhillon}.
The question now is whether such simplifications have a significant impact on network performance,
and whether those conclusions still hold and when.


To address the question with regard to the probability of LOS, 
the authors in~\cite{LOS_NLOS_Trans} have studied the impact of the aforementioned issue
using a path loss model that incorporates both LOS and NLOS transmissions in dense SCNs. 
However, Rayleigh fading was used for both components,
which is not accurate,
as the fading in LOS transmissions are well known to be non-Rayleigh distributed. 
Despite of such simplification,
the authors showed that,  
as the density of small cells increases, 
the ASE will initially increase. 
However, as the density of small cells exceeds a specific threshold, 
the coverage probability will decrease due to the transition of a large number of interfering signals from NLOS transmission to LOS transmission, 
and as a consequent 
the ASE may either continue to grow but at a much slower pace or even decrease. 
In other words, due to such transition, 
the interference power will increase faster than the signal power for some BS densities, 
which negatively affects the user's signal-to-interference-plus-noise-ratio (SINR), 
and verifies that the density of small cell BSs may have a key impact on network performance. 

However, in terms of multi-path fading and as mentioned before,
it is important to note that the authors in~\cite{LOS_NLOS_Trans} oversimplified the LOS link assuming Rayleigh fading. 
An interesting question is whether a more accurate multi-path fading channel model will change the conclusions in~\cite{LOS_NLOS_Trans},
which already changed the conclusions in~\cite{JSAC_Dhillon},
and whether the multi-path fading can mitigate or exacerbate the issues brought by the NLOS to LOS transition. 

In this paper, we consider a Rician fading based channel model, 
which is a more appropriate fading model for dense SCNs, 
especially for LOS paths.
The main contributions of this paper are as follows:
\begin{itemize}
\item 
We derive the analytical results for the coverage probability as well as the ASE of a dense SCN under a Rician fading channel using a general path loss model incorporating both LOS and NLOS transmissions.
\item 
Moreover, we obtain integral-form expressions for the coverage probability and the ASE using a 3GPP path loss model with a \emph{linear} LOS probability function. 
Note that studying the linear LOS probability function not only allows us to obtain more tractable results, 
but also help us to deal with more complicated path loss models in practice, 
as they can be approximated by piece-wise linear functions. 
\item 
Our theoretical analysis reveals an important finding, 
i.e., due to the dominance of paths loss in dense SCNs, 
the impact of the multi-path fading is negligible and thus when the density of small cell BSs exceeds a certain threshold, 
the network coverage probability will decrease as small cells become denser, 
The Rician multi-path fading does not help to mitigate this phenomena. 
\end{itemize}

The remainder of this paper is structured as follows. 
In Section~\ref{sec:System-Model}, 
the system model is presented. 
In Section~\ref{sec:General-Results}, 
the main analytical results on the coverage probability and the ASE taking into account the Rician fading channel are discussed.
In Section~\ref{sec:Simulation-and-Discussion}, 
the numerical results are presented. 
Finally, in Section~\ref{sec:Conclusion}, 
the conclusions are drawn.

\section{System Model\label{sec:System-Model}}

Stochastic geometry is a useful tool to study the performance of the cellular systems. 
In this paper, our focus is on the downlink (DL) of cellular networks. 

\subsubsection*{BS Distribution}
We assume that small cell BSs form a Homogeneous Poisson point process (HPPP) $\Phi$ of intensity $\lambda$ $\textrm{BSs/km}^{2}$.

\subsubsection*{User Distribution}
We assume that UEs form another stationary HPPP with an intensity of $\lambda^{\textrm{UE}}$ $\textrm{UEs/km}^{2}$, 
which is independent from the small cell BSs distribution. 
Note that $\lambda^{\textrm{UE}}$ is considered to be sufficiently larger than $\lambda$ so that each BS has at least one associated UE in its coverage. 
We also assume that a typical UE is located at the origin, 
which is a common assumption in the analysis using stochastic geometry. 

\subsubsection*{Path Loss}
We denote the distance between an arbitrary BS and the typical UE by $r$ in km. 
Considering practical LOS/NLOS transmissions, 
we propose to model the path loss with respect to distance $r$ as shown in (\ref{eq:prop_PL_model}). 
\noindent 
\begin{algorithm*}
\begin{singlespace}
\noindent 
\begin{equation}
\zeta\left(r\right)=\begin{cases}
\zeta_{1}\left(r\right)=\begin{cases}
\begin{array}{l}
\zeta_{1}^{\textrm{L}}\left(r\right),\\
\zeta_{1}^{\textrm{NL}}\left(r\right),
\end{array} & \hspace{-0.3cm}\begin{array}{l}
\textrm{with probability }\textrm{Pr}_{1}^{\textrm{L}}\left(r\right)\\
\textrm{with probability }\left(1-\textrm{Pr}_{1}^{\textrm{L}}\left(r\right)\right)
\end{array}\end{cases}\hspace{-0.3cm}, & \hspace{-0.3cm}\textrm{when }0\leq r\leq d_{1}\\
\zeta_{2}\left(r\right)=\begin{cases}
\begin{array}{l}
\zeta_{2}^{\textrm{L}}\left(r\right),\\
\zeta_{2}^{\textrm{NL}}\left(r\right),
\end{array} & \hspace{-0.3cm}\begin{array}{l}
\textrm{with probability }\textrm{Pr}_{2}^{\textrm{L}}\left(r\right)\\
\textrm{with probability }\left(1-\textrm{Pr}_{2}^{\textrm{L}}\left(r\right)\right)
\end{array}\end{cases}\hspace{-0.3cm}, & \hspace{-0.3cm}\textrm{when }d_{1}<r\leq d_{2}\\
\vdots & \vdots\\
\zeta_{N}\left(r\right)=\begin{cases}
\begin{array}{l}
\zeta_{N}^{\textrm{L}}\left(r\right),\\
\zeta_{N}^{\textrm{NL}}\left(r\right),
\end{array} & \hspace{-0.3cm}\begin{array}{l}
\textrm{with probability }\textrm{Pr}_{N}^{\textrm{L}}\left(r\right)\\
\textrm{with probability }\left(1-\textrm{Pr}_{N}^{\textrm{L}}\left(r\right)\right)
\end{array}\end{cases}\hspace{-0.3cm}, & \hspace{-0.3cm}\textrm{when }r>d_{N-1}
\end{cases}.\label{eq:prop_PL_model}
\end{equation}
\end{singlespace}
\end{algorithm*}

As can be seen from (\ref{eq:prop_PL_model}), 
the path loss function $\zeta\left(r\right)$ is divided into $N$ pieces where each piece is represented by $\zeta_{n}\left(r\right)$. 
Moreover, $\zeta_{n}^{\textrm{L}}\left(r\right)$, $\zeta_{n}^{\textrm{NL}}\left(r\right)$ and $\textrm{Pr}_{n}^{\textrm{L}}\left(r\right)$ represent the $n$-th piece of path loss function for the LOS transmission, the $n$-th piece of path loss function for the NLOS transmission, and the $n$-th piece of the LOS probability function, respectively. 
In addition, we model $\zeta_{n}^{\textrm{L}}\left(r\right)$ and $\zeta_{n}^{\textrm{NL}}\left(r\right)$ in (\ref{eq:prop_PL_model}) as
\noindent 
\begin{equation}
\zeta_{n}\left(r\right)=\begin{cases}
\begin{array}{l}
\zeta_{n}^{\textrm{L}}\left(r\right)=A_{n}^{{\rm {L}}}r^{-\alpha_{n}^{{\rm {L}}}},\\
\zeta_{n}^{\textrm{NL}}\left(r\right)=A_{n}^{{\rm {NL}}}r^{-\alpha_{n}^{{\rm {NL}}}},
\end{array} & \hspace{-0.3cm}\begin{array}{l}
\textrm{for LOS}\\
\textrm{for NLoS}
\end{array},\end{cases}\label{eq:PL_BS2UE}
\end{equation}
where $A_{n}^{{\rm {L}}}$ and $A_{n}^{{\rm {NL}}}, n\in\left\{ 1,2,\ldots,N\right\} $ are the path losses at a reference distance of $r=1~km$ for the LOS and the NLOS cases in $\zeta_{n}\left(r\right)$, respectively, 
and $\alpha_{n}^{{\rm {L}}}$ and $\alpha_{n}^{{\rm {NL}}}, n\in\left\{ 1,2,\ldots,N\right\} $ are the path loss exponents for the LOS and the NLOS cases in $\zeta_{n}\left(r\right)$, respectively. 
Typical values of reference path losses and path loss exponents can be found in~\cite{TR36.828},~\cite{SCM_pathloss_model}.
Furthermore, we stack $\left\{ \zeta_{n}^{\textrm{L}}\left(r\right)\right\} $ and $\left\{ \zeta_{n}^{\textrm{NL}}\left(r\right)\right\} $ into piece-wise functions as
\begin{singlespace}
\noindent 
\begin{equation}
\zeta^{s}\left(r\right)=\begin{cases}
\zeta_{1}^{s}\left(r\right), & \textrm{when }0\leq r\leq d_{1}\\
\zeta_{2}^{s}\left(r\right),\hspace{-0.3cm} & \textrm{when }d_{1}<r\leq d_{2}\\
\vdots & \vdots\\
\zeta_{N}^{s}\left(r\right), & \textrm{when }r>d_{N-1}
\end{cases},\label{eq:general_PL_func}
\end{equation}
\end{singlespace}
\noindent where $s$ is a string variable taking the values from $s \in \{\rm{L},\rm{NL}\}$, for the LOS and the NLOS cases, respectively. 

Note that in (\ref{eq:prop_PL_model}), 
$\textrm{Pr}_{n}^{\textrm{L}}\left(r\right),n\in\left\{ 1,2,\ldots,N\right\} $ denotes the $n$-th piece LOS probability function corresponding to a BS and a UE that are separated by the distance $r$.

\subsubsection*{User Association Strategy (UAS)}
The UE is associated with the BS with the smallest path loss, 
regardless whether it is LOS or NLOS.

\subsubsection*{Antenna Radiation Pattern}
Each BS and the typical UE are equipped with an isotropic antenna.

\subsubsection*{Multi-path Fading}
The multi-path fading between an arbitrary BS and the typical UE is modelled as distance dependant Rician fading channel. 
More specifically, the Rician K factor is defined as the ratio of the power in the specular LOS component to the power in all NLOS components. 
For the LOS case, we use a distance dependant Rician K factor where $K = 13 - 0.03r \ (dB)$ 
and $r$ denotes the distance between the BS and UE in meter~\cite{TR25.996}. 
For the NLOS case, the Rician K factor is set to $0 \ dB$.

\section{Analysis Based on the Proposed Path Loss Model \label{sec:General-Results}}

The coverage probability represents the probability that the UE's signal-to-interference-plus-noise-ratio (SINR) is above a threshold $\gamma$:
\begin{singlespace}
\noindent 
\begin{equation}
p^{\textrm{cov}}\left(\lambda,\gamma\right)=\textrm{Pr}\left[\mathrm{SINR}>\gamma\right],\label{eq:Coverage_Prob_def}
\end{equation}
\end{singlespace}
\noindent where the SINR is computed by
\noindent 
\begin{equation}
\mathrm{SINR}=\frac{P\zeta\left(r\right)h}{I_{r}+N_{0}},\label{eq:SINR}
\end{equation}
where $h$ is the Rician distributed channel gain, and $P$ and $N_{0}$ refer to the transmission power of each BS and the additive white Gaussian noise (AWGN) power at the typical UE, respectively.  
The aggregated interference from all non-serving BSs is denoted by $I_{r}$,  
and is defined as
\noindent 
\begin{equation}
I_{r}=\sum_{i:\, b_{i}\in\Phi\setminus b_{o}}P\beta_{i}g_{i},\label{eq:cumulative_interference}
\end{equation}
where $b_{o}$ denotes the serving BS, and  $b_{i}$, $\beta_{i}$ and $g_{i}$ refer to the $i$-th interfering BS, the corresponding path loss of $b_{i}$ and the corresponding Rician fading channel gain of $b_{i}$, respectively.

For a specific $\lambda$, the area spectral efficiency (ASE) in $\textrm{bps/Hz/km}^{2}$  can be expressed as 
\begin{singlespace}
\noindent 
\begin{equation}
A^{\textrm{ASE}}\left(\lambda,\gamma_{0}\right)=\lambda\int_{\gamma_{0}}^{\infty}\log_{2}\left(1+\gamma\right)f_{\mathit{\Gamma}}\left(\lambda,\gamma\right)d\gamma,\label{eq:ASE_def}
\end{equation}
\end{singlespace}
\noindent where $\gamma_{0}$ denotes the minimum working SINR for the considered SCN, 
and $f_{\mathit{\Gamma}}\left(\lambda,\gamma\right)$ represents the probability density function (PDF) of SINR for a specific value of $\lambda$ at the typical UE. 
Note that the ASE defined in this paper is different from that in \cite{Jeff's work 2011}, 
where a deterministic rate based on $\gamma_0$ is assumed for the typical UE, 
no matter what the actual SINR value is. 
The ASE definition in (\ref{eq:ASE_def}) is more realistic due to the SINR-dependent rate, 
but it is more complex to analyse, 
as it requires one more fold of numerical integral compared with~\cite{Jeff's work 2011}.

The PDF of SINR is then computed as
\begin{singlespace}
\noindent 
\begin{equation}
f_{\mathit{\Gamma}}\left(\lambda,\gamma\right)=\frac{\partial\left(1-p^{\textrm{cov}}\left(\lambda,\gamma\right)\right)}{\partial\gamma}.\label{eq:cond_SINR_PDF}
\end{equation}
\end{singlespace}
 
In the following, we present Theorem~\ref{thm:p_cov_UAS1}, 
which is used to obtain the $p^{\textrm{cov}}\left(\lambda,\gamma\right)$ based on the proposed path loss model in (\ref{eq:prop_PL_model}). 
Note that for tractability of analysis, we consider an interference limited scenario where $I_{r} \gg N_{0}$.

\begin{thm}
\label{thm:p_cov_UAS1}
Considering the path loss model in (\ref{eq:prop_PL_model}), 
$p^{\textrm{cov}}\left(\lambda,\gamma\right)$ is computed as
\begin{equation}
p^{\textrm{cov}}\left(\lambda,\gamma\right)=\sum_{n=1}^{N}\left(T_{n}^{\textrm{L}}+T_{n}^{\textrm{NL}}\right),\label{eq:Theorem_1_p_cov}
\end{equation}
where $T_{n}^{\textrm{L}}=\int_{d_{n-1}}^{d_{n}}\textrm{Pr}\left[\frac{P\zeta_{n}^{\textrm{L}}\left(r\right)h}{I_{r}}>\gamma\right]f_{R,n}^{\textrm{L}}\left(r\right)dr$,
$T_{n}^{\textrm{NL}}=\int_{d_{n-1}}^{d_{n}}\textrm{Pr}\left[\frac{P\zeta_{n}^{\textrm{NL}}\left(r\right)h}{I_{r}}>\gamma\right]f_{R,n}^{\textrm{NL}}\left(r\right)dr$,
and $d_{0}=0$ and $d_{N}=\infty$. 

Moreover, $f_{R,n}^{\textrm{L}}\left(r\right)$ and $f_{R,n}^{\textrm{NL}}\left(r\right)$ are defined by
\begin{eqnarray}
f_{R,n}^{\textrm{L}}\left(r\right) & = & \exp\left(-\int_{0}^{r_{1}}\left(1-\textrm{Pr}^{\textrm{L}}\left(u\right)\right)2\pi u\lambda du\right)\times\exp\left(-\int_{0}^{r}\textrm{Pr}^{\textrm{L}}\left(u\right)2\pi u\lambda du\right)\nonumber \\
 &  & \times\textrm{Pr}_{n}^{\textrm{L}}\left(r\right)\times2\pi r\lambda,\quad\left(d_{n-1}<r\leq d_{n}\right),\label{eq:geom_dis_PDF_UAS1_LoS_thm}
\end{eqnarray}
and
\begin{eqnarray}
f_{R,n}^{\textrm{NL}}\left(r\right) & = & \exp\left(-\int_{0}^{r_{2}}\textrm{Pr}^{\textrm{L}}\left(u\right)2\pi u\lambda du\right)\times\exp\left(-\int_{0}^{r}\left(1-\textrm{Pr}^{\textrm{L}}\left(u\right)\right)2\pi u\lambda du\right)\nonumber \\
 &  & \times\left(1-\textrm{Pr}_{n}^{\textrm{L}}\left(r\right)\right)\times2\pi r\lambda,\quad\left(d_{n-1}<r\leq d_{n}\right),\label{eq:geom_dis_PDF_UAS1_NLoS_thm}
\end{eqnarray}
It is noteworthy that we can determine $r_{1}$ and $r_{2}$ as $\underset{r_{1}}{\arg}\left\{ \zeta^{\textrm{NL}}\left(r_{1}\right)=\zeta_{n}^{\textrm{L}}\left(r\right)\right\} $ and $\underset{r_{2}}{\arg}\left\{ \zeta^{\textrm{L}}\left(r_{2}\right)=\zeta_{n}^{\textrm{NL}}\left(r\right)\right\} $, respectively.

Furthermore, $\textrm{Pr}\left[\frac{P\zeta_{n}^{\textrm{L}}\left(r\right)h}{I_{r}}>\gamma\right]$
and $\textrm{Pr}\left[\frac{P\zeta_{n}^{\textrm{NL}}\left(r\right)h}{I_{r}}>\gamma\right]$
are respectively computed by 
\noindent 
\begin{equation}
{\rm {Pr}}[\frac{P \zeta_{n}^{L}(r)h}{I_{r}}>\gamma] = \sum\limits_{k=0}^{\infty} \sum\limits_{m = 0}^{k} \ J(m,k) \ \gamma^{k-m}(-1)^{k-m} \frac{\partial^{k-m} \mathscr{L}_{I_{r}}(\frac{\gamma}{P \zeta_{n}^{L}(r)})}{\partial \gamma^{k-m}}
\label{eq:Pr_SINR_req_UAS1_LoS_thm}
\end{equation}
and
\begin{equation}
{\rm {Pr}}[\frac{P \zeta_{n}^{NL}(r)h}{I_{r}}>\gamma] = \sum\limits _{k=0}^{\infty} \sum\limits_{m = 0}^{k} \ J(m,k) \ \gamma^{k-m}(-1)^{k-m} \frac{\partial^{k-m} \mathscr{L}_{I_{r}}(\frac{\gamma}{P \zeta_{n}^{NL}(r)})}{\partial \gamma^{k-m}}
\label{eq:Pr_SINR_req_UAS1_NLoS_thm}
\end{equation}
where $\mathscr{L}_{I_{r}}\left(s\right)$ is the Laplace transform of RV $I_{r}$ evaluated at $s$.\end{thm}

\begin{IEEEproof}
See Appendix~A. 
\end{IEEEproof}

\section{Study of a 3GPP Special Case \label{sec:A-3GPP-Special-Case}}

We consider the following path loss function, $\zeta\left(r\right)$, which is adopted by the 3GPP in\cite{TR36.828}
\noindent 
\begin{equation}
\zeta\left(r\right)=\begin{cases}
\begin{array}{l}
A^{{\rm {L}}}r^{-\alpha^{{\rm {L}}}},\\
A^{{\rm {NL}}}r^{-\alpha^{{\rm {NL}}}},
\end{array}\hspace{-0.3cm} & \begin{array}{l}
\textrm{with probability }\textrm{Pr}^{\textrm{L}}\left(r\right)\\
\textrm{with probability }\left(1-\textrm{Pr}^{\textrm{L}}\left(r\right)\right)
\end{array}\end{cases},\label{eq:PL_BS2UE_2slopes}
\end{equation}
which for sake of simplicity and without any loss of generality uses a linear LOS probability\cite{SCM_pathloss_model} function, $\textrm{Pr}^{\textrm{L}}\left(r\right)$, defined as
\begin{singlespace}
\noindent 
\begin{equation}
\textrm{Pr}^{\textrm{L}}\left(r\right)=\begin{cases}
\begin{array}{l}
1-\frac{r}{d_{1}},\\
0,
\end{array}\hspace{-0.3cm} & \begin{array}{l}
0<r\leq d_{1}\\
r>d_{1}
\end{array}\end{cases},\label{eq:LoS_Prob_func_linear}
\end{equation}
\end{singlespace}
\noindent where the steepness of $\textrm{Pr}^{\textrm{L}}\left(r\right)$ is defined by the parameter $d_{1}$.

According to Theorem~\ref{thm:p_cov_UAS1} and considering the mentioned 3GPP case, 
$p^{\textrm{cov}}\left(\lambda,\gamma\right)$ can be computed as $p^{\textrm{cov}}\left(\lambda,\gamma\right)=\sum\limits_{n=1}^{2}\left(T_{n}^{\textrm{L}}+T_{n}^{\textrm{NL}}\right)$, 
and hence in the following, 
we present $T_{1}^{\textrm{L}}$, $T_{1}^{\textrm{NL}}$, $T_{2}^{\textrm{L}}$, and $T_{2}^{\textrm{NL}}$, respectively. 

\subsection{The Computation of $T_{1}^{\textrm{L}}$\label{sub:The-Computation-of-T1L}}

From Theorem~\ref{thm:p_cov_UAS1}, 
$T_{1}^{\textrm{L}}$ is computed as 
\noindent 
\begin{eqnarray}
T_{1}^{\textrm{L}} & = & \int_{0}^{d_{1}}\sum\limits_{k=0}^{\infty} \sum\limits_{m = 0}^{k} \ J(m,k) \ \gamma^{k-m}(-1)^{k-m} \frac{\partial^{k-m} \mathscr{L}_{I_{r}}(\frac{\gamma}{P \zeta_{n}^{L}(r)})}{\partial \gamma^{k-m}} f_{R,1}^{\textrm{L}}\left(r\right)dr\nonumber \\
& \overset{(a)}{=} & \int_{0}^{d_{1}} \sum\limits_{k=0}^{\infty} \sum\limits_{m = 0}^{k} \ J(m,k) \ \gamma^{k-m}(-1)^{k-m} \frac{\partial^{k-m} \mathscr{L}_{I_{r}}(\frac{\gamma r^{\alpha^{{\rm {L}}}}}{PA^{{\rm {L}}}})}{\partial \gamma^{k-m}} f_{R,1}^{\textrm{L}}\left(r\right)dr,\label{eq:T_1_UAS1_LoS}
\end{eqnarray}
where $\zeta_{1}^{\textrm{L}}\left(r\right)=A^{{\rm {L}}}r^{-\alpha^{{\rm {L}}}}$ from (\ref{eq:PL_BS2UE_2slopes}) is plugged into the step (a) of (\ref{eq:T_1_UAS1_LoS}). 
Note that $\mathscr{L}_{I_{r}}\left(s\right)$ represents the Laplace transform of RV $I_{r}$ evaluated at $s$. 

In (\ref{eq:T_1_UAS1_LoS}), 
according to Theorem~\ref{thm:p_cov_UAS1} and (\ref{eq:LoS_Prob_func_linear}), 
$f_{R,1}^{\textrm{L}}\left(r\right)$ is computed as
\noindent 
\begin{eqnarray}
f_{R,1}^{\textrm{L}}\left(r\right) & = & \exp\left(-\int_{0}^{r_{1}}\lambda\frac{u}{d_{1}}2\pi udu\right)\exp\left(-\int_{0}^{r}\lambda\left(1-\frac{u}{d_{1}}\right)2\pi udu\right)\left(1-\frac{r}{d_{1}}\right)2\pi r\lambda\nonumber \\
 & = & \exp\left(-\pi\lambda r^{2}+2\pi\lambda\left(\frac{r^{3}}{3d_{1}}-\frac{r_{1}^{3}}{3d_{1}}\right)\right)\left(1-\frac{r}{d_{1}}\right)2\pi r\lambda,\quad\left(0<r\leq d_{1}\right),\label{eq:spec_geom_dis_PDF_UAS1_LoS_seg1}
\end{eqnarray}
where $r_{1}=\left(\frac{A^{{\rm {NL}}}}{A^{{\rm {L}}}}\right)^{\frac{1}{\alpha^{{\rm {NL}}}}}r^{\frac{\alpha^{{\rm {L}}}}{\alpha^{{\rm {NL}}}}}$. 
Moreover, in order to compute $\mathscr{L}_{I_{r}}\left(\frac{\gamma r^{\alpha^{{\rm {L}}}}}{PA^{{\rm {L}}}}\right)$ in (\ref{eq:T_1_UAS1_LoS}) for the range of $0<r\leq d_{1}$, 
we propose Lemma~\ref{lem:laplace_term_UAS1_LoS_seg1}. 
\begin{lem}

\noindent \label{lem:laplace_term_UAS1_LoS_seg1}
$\mathscr{L}_{I_{r}}\left(\frac{\gamma r^{\alpha^{{\rm {L}}}}}{PA^{{\rm {L}}}}\right)$ in the range of $0<r\leq d_{1}$ can be computed by
\noindent
 
\begin{equation}
\begin{aligned}
\hspace{-0.1cm} \mathscr{L}_{I_{r}}\left(\frac{\gamma r^{\alpha^{{\rm {L}}}}}{PA^{{\rm {L}}}}\right) = 
\\ & \hspace{-2.5cm} {\rm exp}\left(-2\pi\lambda \ \left(\rho_{1}\left(\alpha^{L},1,(1-K)\left(\gamma r^{\alpha^{\rm L}}
\right)^{-1},d_{1}\right) \ - \rho_{1}\left(\alpha^{L},1,(1-K)\left(\gamma r^{\alpha^{\rm L}}\right)^{-1},r\right)\right)\right)
\\ & \hspace{-2.7cm} \times \ {\rm exp}\biggl(-2\pi\lambda \ \biggl(\rho_{1}\left(\alpha^{L},\alpha^{L} + 1,(1-K)\left(\gamma r^{\alpha^{\rm L}}\right)^{-1},d_{1}\right) 
\\ & \hspace{-2.7cm} - \ \rho_{1}\left(\alpha^{L},\alpha^{L}+ 1,(1-K)\left(\gamma r^{\alpha^{\rm L}}\right)^{-1},r\right)\biggr)\biggr)
\\ & \hspace{-2.7cm} \times \ {\rm exp}\biggl(\frac{2\pi\lambda}{d_{0}}\left(\gamma r^{\alpha^{\rm L}}\right)^{-1} \left(1-K-e^{-K}\right)\biggl(\rho_{1}\left(\alpha^{L},2,(1-K)\left(\gamma r^{\alpha^{\rm L}}\right)^{-1},d_{1}\right)
\\ & \hspace{-2.7cm} - \ \rho_{1}\left(\alpha^{L},2,(1-K)\left(\gamma r^{\alpha^{\rm L}}\right)^{-1},r\right)\biggr)\biggr)
\\ & \hspace{-2.7cm} \times \ {\rm exp}\biggl(\frac{2\pi\lambda}{d_{0}}\left(\gamma r^{\alpha^{\rm L}}\right)^{-1} \left(1-K-e^{-K}\right) \biggl(\rho_{1}\left(\alpha^{L},\alpha^{L}+2,(1-K)\left(\gamma r^{\alpha^{\rm L}}\right)^{-1},d_{1}\right) 
\\ & \hspace{-2.7cm} - \ \rho_{1}\left(\alpha^{L},\alpha^{L}+2,(1-K)\left(\gamma r^{\alpha^{\rm L}}\right)^{-1},r_{1}\right)\biggr)\biggr)
\\ & \hspace{-2.7cm} \times \ {\rm exp}\Biggl(\frac{-2\pi\lambda}{d_{0}} \Biggl(\rho_{1}\left(\alpha^{NL},2,(1-K)\left(\frac{\gamma A^{\rm NL}}{A^{\rm L}} r^{\alpha^{\rm L}}\right)^{-1},d_{1}\right)
\\ & \hspace{-2.7cm} - \ \rho_{1}\left(\alpha^{NL},2,(1-K)\left(\frac{\gamma A^{\rm NL}}{A^{\rm L}} r^{\alpha^{\rm L}}\right)^{-1},r_{1}\right)\Biggr)\Biggr)
\\ & \hspace{-2.7cm} \times \ {\rm exp}\Biggl(\frac{-2\pi\lambda}{d_{0}}\left(\frac{\gamma A^{\rm NL}}{A^{\rm L}}r^{\alpha^{\rm L}}\right)^{-1}\left(1-K-e^{-K}\right) \Biggl(\rho_{1}\left(\alpha^{NL},\alpha^{NL}+ 2,(1-K)\left(\frac{\gamma A^{\rm NL}}{A^{\rm L}} r^{\alpha^{\rm L}}\right)^{-1},d_{1}\right)
\\ & \hspace{-2.7cm} - \ \rho_{1}\left(\alpha^{NL},\alpha^{NL}+ 2,(1-K)\left(\frac{\gamma A^{\rm NL}}{A^{\rm L}} r^{\alpha^{\rm L}}\right)^{-1},r_{1}\right)\Biggr)\Biggr)
\\ & \hspace{-2.7cm} \times \ {\rm exp}\left(-2\pi\lambda \rho_{2}\left(\alpha^{NL},1,(1-K)\left(\frac{\gamma A^{\rm NL}}{A^{\rm L}} r^{\alpha^{\rm L}}\right)^{-1},d_{1}\right)\right)
\\ & \hspace{-2.7cm} \times \ {\rm exp}\Biggl(-2\pi\lambda \left(\frac{\gamma A^{\rm NL}}{A^{\rm L}} r^{\alpha^{\rm L}}\right)^{-1} \left(1-K-e^{-K}\right) 
\\ & \hspace{-2.5cm} \rho_{2}\left(\alpha^{NL},\alpha^{NL}+1,(1-K)\left(\frac{\gamma A^{\rm NL}}{A^{\rm L}} r^{\alpha^{\rm L}}\right)^{-1},d_{1}\right)\Biggr), \quad \quad (0<r \leq d_{1})
\label{eq:Lemma_3}
\end{aligned}
\end{equation}
where 
\begin{equation}
\rho_{1}\left(\alpha,\beta,t,d\right)=\left[\frac{d^{\left(\beta+1\right)}}{\beta+1}\right]{}_{2}F_{1}\left[1,\frac{\beta+1}{\alpha};1+\frac{\beta+1}{\alpha};-td^{\alpha}\right],\label{eq:rou1_func}
\end{equation}
and
\begin{equation}
\rho_{2}\left(\alpha,\beta,t,d\right)=\left[\frac{d^{-\left(\alpha-\beta-1\right)}}{t\left(\alpha-\beta-1\right)}\right]{}_{2}F_{1}\left[1,1-\frac{\beta+1}{\alpha};2-\frac{\beta+1}{\alpha};-\frac{1}{td^{\alpha}}\right],\left(\alpha>\beta+1\right),\label{eq:rou2_func}
\end{equation}
where $_{2}F_{1}\left[\cdot,\cdot;\cdot;\cdot\right]$ is the hyper-geometric function~\cite{Book_Integrals}.
\end{lem}

\begin{IEEEproof}
See Appendix~B. 
\end{IEEEproof}
Overall, we can evaluate $T_{1}^{\textrm{L}}$ as
\begin{equation}
T_{1}^{\textrm{L}}=\int_{0}^{d_{1}} \sum\limits_{k=0}^{\infty} \sum\limits_{m = 0}^{k} \ J(m,k) \ \gamma^{k-m}(-1)^{k-m} \frac{\partial^{k-m} \mathcal{L}_{I_{r}}(\frac{\gamma r^{\alpha^{{\rm {L}}}}}{PA^{{\rm {L}}}})}{\partial \gamma^{k-m}} f_{R,1}^{\textrm{L}}\left(r\right)dr,\label{eq:T_1_UAS1_LoS_final}
\end{equation}
where $f_{R,1}^{\textrm{L}}\left(r\right)$ and $\mathscr{L}_{I_{r}}\left(\frac{\gamma r^{\alpha^{{\rm {L}}}}}{PA^{{\rm {L}}}}\right)$ are determined by (\ref{eq:spec_geom_dis_PDF_UAS1_LoS_seg1}) and (\ref{eq:Lemma_3}), respectively.

\subsection{The Computation of $T_{1}^{\textrm{NL}}$\label{sub:The-Computation-of-T1NL}}

From Theorem~\ref{thm:p_cov_UAS1}, 
$T_{1}^{\textrm{NL}}$ is computed as 
\begin{eqnarray}
T_{1}^{\textrm{NL}} & = & \int_{0}^{d_{1}}\sum\limits_{k=0}^{\infty} \sum\limits_{m=0}^{k} \ J(m,k) \ \gamma^{k-m}(-1)^{k-m} \frac{\partial^{k-m} \mathscr{L}_{I_{r}}(\frac{\gamma}{P \zeta_{n}^{NL}(r)})}{\partial \gamma^{k-m}} f_{R,1}^{\textrm{NL}}\left(r\right)dr \nonumber\\
& \overset{(a)}{=} & \int_{0}^{d_{1}} \sum\limits_{k=0}^{\infty} \sum\limits_{m = 0}^{k} \ J(m,k) \ \gamma^{k-m}(-1)^{k-m} \frac{\partial^{k-m} \mathscr{L}_{I_{r}}(\frac{\gamma r^{\alpha^{{\rm {NL}}}}}{PA^{{\rm {NL}}}})}{\partial \gamma^{k-m}} f_{R,1}^{\textrm{NL}}\left(r\right)dr,\label{eq:T_1_UAS1_NLoS}
\end{eqnarray}
where $\zeta_{1}^{\textrm{NL}}\left(r\right)=A^{{\rm {NL}}}r^{-\alpha^{{\rm {NL}}}}$ from (\ref{eq:PL_BS2UE_2slopes}) is plugged into the step (a) of (\ref{eq:T_1_UAS1_NLoS}). 

In (\ref{eq:T_1_UAS1_NLoS}), 
according to Theorem~\ref{thm:p_cov_UAS1} and (\ref{eq:LoS_Prob_func_linear}), 
$f_{R,1}^{\textrm{NL}}\left(r\right)$ can be written as
\begin{eqnarray}
f_{R,1}^{\textrm{NL}}\left(r\right) & = & \exp\left(-\int_{0}^{r_{2}}\lambda\textrm{Pr}^{\textrm{L}}\left(u\right)2\pi udu\right)\nonumber \\
 &  & \times\exp\left(-\int_{0}^{r}\lambda\left(1-\textrm{Pr}^{\textrm{L}}\left(u\right)\right)2\pi udu\right)\left(\frac{r}{d_{1}}\right)2\pi r\lambda,\quad\left(0<r\leq d_{1}\right),\label{eq:spec_geom_dis_PDF_UAS1_NLoS_seg1}
\end{eqnarray}
where $r_{2}=\left(\frac{A^{{\rm {L}}}}{A^{{\rm {NL}}}}\right)^{\frac{1}{\alpha^{{\rm {L}}}}}r^{\frac{\alpha^{{\rm {NL}}}}{\alpha^{{\rm {L}}}}}$. 
In the following, we discuss the cases of $0<r_{2}\leq d_{1}$ and $r_{2}>d_{1}$ separately.

If $0<r_{2}\leq d_{1}$, i.e., $0<r\leq y_{1}=d_{1}^{\frac{\alpha^{{\rm {L}}}}{\alpha^{{\rm {NL}}}}}\left(\frac{A^{{\rm {NL}}}}{A^{{\rm {L}}}}\right)^{\frac{1}{\alpha^{{\rm {NL}}}}}$, 
the $f_{R,1}^{\textrm{NL}}\left(r\right)$ is calculated as
\begin{eqnarray}
f_{R,1}^{\textrm{NL}}\left(r\right)\hspace{-0.3cm} & = & \hspace{-0.3cm}\exp\left(-\int_{0}^{r_{2}}\lambda\left(1-\frac{u}{d_{1}}\right)2\pi udu\right)\exp\left(-\int_{0}^{r}\lambda\frac{u}{d_{1}}2\pi udu\right)\left(\frac{r}{d_{1}}\right)2\pi r\lambda\nonumber \\
\hspace{-0.3cm} & = & \hspace{-0.3cm}\exp\left(-\pi\lambda r_{2}^{2}+2\pi\lambda\left(\frac{r_{2}^{3}}{3d_{1}}-\frac{r^{3}}{3d_{1}}\right)\right)\left(\frac{r}{d_{1}}\right)2\pi r\lambda,\quad\hspace{-0.3cm}\left(0<r\leq y_{1}\right).\label{eq:spec_geom_dis_PDF_UAS1_NLoS_seg1_case1}
\end{eqnarray}
Otherwise, if $r_{2}>d_{1}$, i.e., $y_{1}<r\leq d_{1}$, 
the $f_{R,1}^{\textrm{NL}}\left(r\right)$ is calculated as
\begin{eqnarray}
f_{R,1}^{\textrm{NL}}\left(r\right)\hspace{-0.3cm} & = & \hspace{-0.3cm}\exp\left(-\int_{0}^{d_{1}}\lambda\left(1-\frac{u}{d_{1}}\right)2\pi udu\right)\exp\left(-\int_{0}^{r}\lambda\frac{u}{d_{1}}2\pi udu\right)\left(\frac{r}{d_{1}}\right)2\pi r\lambda\nonumber \\
\hspace{-0.3cm} & = & \hspace{-0.3cm}\exp\left(-\frac{\pi\lambda d_{1}^{2}}{3}-\frac{2\pi\lambda r^{3}}{3d_{1}}\right)\left(\frac{r}{d_{1}}\right)2\pi r\lambda,\quad\left(y_{1}<r\leq d_{1}\right).\label{eq:spec_geom_dis_PDF_UAS1_NLoS_seg1_case2}
\end{eqnarray}

In the following, Lemma~\ref{lem:laplace_term_UAS1_NLoS_seg1} is proposed to compute $\mathscr{L}_{I_{r}}\left(\frac{\gamma r^{\alpha^{{\rm {NL}}}}}{PA^{{\rm {NL}}}}\right)$ in (\ref{eq:T_1_UAS1_NLoS}) for the range of $0<r\leq d_{1}$. 
Note that, the computation of $\mathscr{L}_{I_{r}}\left(\frac{\gamma r^{\alpha^{{\rm {NL}}}}}{PA^{{\rm {NL}}}}\right)$ will also be performed separately in the two ranges of $0<r_{2}\leq d_{1}$ and $r_{2}>d_{1}$. 

\begin{lem}
\noindent \label{lem:laplace_term_UAS1_NLoS_seg1}$\mathscr{L}_{I_{r}}\left(\frac{\gamma r^{\alpha^{{\rm {NL}}}}}{PA^{{\rm {NL}}}}\right)$ in the range of $0<r\leq d_{1}$ is considered separately for two different cases, i.e., $0<r\leq y_{1}$ and $y_{1}<r\leq d_{1}$. 

\begin{equation}
\begin{aligned}
\hspace{-0.3cm} \mathscr{L}_{I_{r}}\left(\frac{\gamma r^{\alpha^{{\rm {NL}}}}}{PA^{{\rm {NL}}}}\right)\hspace{-0.1cm} = 
\\ & \hspace{-2.7cm} {\rm exp}\Biggl(-2\pi\lambda \ \Biggl(\rho_{1}\left(\alpha^{L},1,(1-K)\left(\frac{\gamma A^{\rm L}}{A^{\rm NL}} r^{\alpha^{\rm NL}}\right)^{-1},d_{1}\right)
\\ & \hspace{-2.7cm} - \rho_{1}\left(\alpha^{L},1,(1-K)\left(\frac{\gamma A^{\rm L}}{A^{\rm NL}} r^{\alpha^{\rm NL}}\right)^{-1},r_{2}\right)\Biggr)\Biggr)
\\ & \hspace{-2.7cm} \times \ {\rm exp}\Biggl(-2\pi\lambda \ \Biggl(\rho_{1}\left(\alpha^{L},\alpha^{L} + 1,(1-K)\left(\frac{\gamma A^{\rm L}}{A^{\rm NL}} r^{\alpha^{\rm NL}}\right)^{-1},d_{1}\right) 
\\ & \hspace{-2.7cm} - \ \rho_{1}\left(\alpha^{L},\alpha^{L}+ 1,(1-K)\left(\frac{\gamma A^{\rm L}}{A^{\rm NL}} r^{\alpha^{\rm NL}}\right)^{-1},r_{2}\right)\Biggr)\Biggr)
\\ & \hspace{-2.7cm} \times \ {\rm exp}\Biggl(\frac{2\pi\lambda}{d_{0}}\left(\frac{\gamma A^{\rm L}}{A^{\rm NL}} r^{\alpha^{\rm NL}}\right)^{-1} \left(1-K-e^{-K}\right)\Biggl(\rho_{1}\left(\alpha^{L},2,(1-K)\left(\frac{\gamma A^{\rm L}}{A^{\rm NL}} r^{\alpha^{\rm NL}}\right)^{-1},d_{1}\right)
\\ & \hspace{-2.7cm} - \ \rho_{1}\left(\alpha^{L},2,(1-K)\left(\frac{\gamma A^{\rm L}}{A^{\rm NL}} r^{\alpha^{\rm NL}}\right)^{-1},r_{2}\right)\Biggr)\Biggr)
\\ & \hspace{-2.7cm} \times \ {\rm exp}\Biggl(\frac{2\pi\lambda}{d_{0}}\left(\frac{\gamma A^{\rm L}}{A^{\rm NL}} r^{\alpha^{\rm NL}}\right)^{-1} \left(1-K-e^{-K}\right) \Biggl(\rho_{1}\left(\alpha^{L},\alpha^{L}+2,(1-K)\left(\frac{\gamma A^{\rm L}}{A^{\rm NL}} r^{\alpha^{\rm NL}}\right)^{-1},d_{1}\right) 
\\ & \hspace{-2.7cm} - \ \rho_{1}\left(\alpha^{L},\alpha^{L}+2,(1-K)\left(\frac{\gamma A^{\rm L}}{A^{\rm NL}} r^{\alpha^{\rm NL}}\right)^{-1},r_{2}\right)\Biggr)\Biggr)
\\ & \hspace{-2.7cm} \times {\rm exp}\biggl(\frac{-2\pi\lambda}{d_{0}} \biggl(\rho_{1}\left(\alpha^{NL},2,(1-K)\left(\gamma r^{\alpha^{\rm NL}}\right)^{-1},d_{1}\right) - \ \rho_{1}\left(\alpha^{NL},2,(1-K)\left(\gamma r^{\alpha^{\rm NL}}\right)^{-1},r\right)\biggr)\biggr)
\\ & \hspace{-2.7cm} \times \ {\rm exp}\biggr(\frac{-2\pi\lambda}{d_{0}}\left(\gamma r^{\alpha^{\rm NL}}\right)^{-1}\left(1-K-e^{-K}\right) \biggl(\rho_{1}\left(\alpha^{NL},\alpha^{NL}+ 2,(1-K)\left(\gamma r^{\alpha^{\rm NL}}\right)^{-1},d_{1}\right)
\\ & \hspace{-2.7cm} - \ \rho_{1}\left(\alpha^{NL},\alpha^{NL}+ 2,(1-K)\left(\gamma r^{\alpha^{\rm NL}}\right)^{-1},r\right)\biggr)\biggr) 
\\ & \hspace{-2.7cm} \times \ {\rm exp}\left(-2\pi\lambda \ \rho_{2}\left(\alpha^{NL},1,(1-K)\left(\gamma r^{\alpha^{\rm NL}}\right)^{-1},d_{1}\right)\right) \times \ {\rm exp}\biggl(-2\pi\lambda \left(\gamma r^{\alpha^{\rm NL}}\right)^{-1} \left(1-K-e^{-K}\right) 
\\ & \hspace{-2.4cm} \rho_{2}\left(\alpha^{NL},\alpha^{NL}+1,(1-K)\left(\gamma r^{\alpha^{\rm NL}}\right)^{-1},d_{1}\right)\biggr) \quad \quad \left(0<r\leq y_{1}\right),
\label{eq:Lemma_4-1}
\end{aligned}
\end{equation}
\vspace{-0.2cm}
and
\vspace{-0.2cm}
\begin{equation}
\begin{aligned}
\mathscr{L}_{I_{r}}\left(\frac{\gamma r^{\alpha^{{\rm {NL}}}}}{PA^{{\rm {NL}}}}\right) = 
\\ & \hspace{-2.8cm} {\rm exp}\left(\frac{-2\pi\lambda}{d_{0}} \left(\rho_{1}\left(\alpha^{NL},2,(1-K)\left(\gamma r^{\alpha^{\rm NL}}\right)^{-1},d_{1}\right) \ - \rho_{1}\left(\alpha^{NL},2,(1-K)\left(\gamma r^{\alpha^{\rm NL}}\right)^{-1},r\right)\right)\right) \\
& \hspace{-2.9cm} \times \ {\rm exp}\Biggl(\frac{-2\pi\lambda}{d_{0}}\left(\gamma r^{\alpha^{\rm NL}}\right)^{-1}\left(1-K-e^{-K}\right) \biggl(\rho_{1}\left(\alpha^{NL},\alpha^{NL}+ 2,(1-K)\left(\gamma r^{\alpha^{\rm NL}}\right)^{-1},d_{1}\right) \\
& \hspace{-2.9cm} - \ \rho_{1}\left(\alpha^{NL},\alpha^{NL}+ 2,(1-K)\left(\gamma r^{\alpha^{\rm NL}}\right)^{-1},r\right)\biggr)\Biggr) \\ & \hspace{-2.9cm} \times \ {\rm exp}\biggl(-2\pi\lambda \ \rho_{2}\left(\alpha^{NL},1,(1-K)\left(\gamma r^{\alpha^{\rm NL}}\right)^{-1},d_{1}\right)\biggr)
\\ & \hspace{-2.9cm} \times \ {\rm exp}\biggl(-2\pi\lambda \left(\gamma r^{\alpha^{\rm NL}}\right)^{-1} \left(1-K-e^{-K}\right) \\
& \hspace{-2.7cm} \rho_{2}\left(\alpha^{NL},\alpha^{NL}+1,(1-K)\left(\gamma r^{\alpha^{\rm NL}}\right)^{-1},d_{1}\right)\biggr) \quad \quad \left(y_{1}<r\leq d_{1}\right),
\label{eq:Lemma_4-2}
\end{aligned}
\end{equation}
where $\rho_{1}\left(\alpha,\beta,t,d\right)$ and $\rho_{2}\left(\alpha,\beta,t,d\right)$ are defined in (\ref{eq:rou1_func}) and (\ref{eq:rou2_func}), respectively. 
\end{lem}

\begin{IEEEproof}
See Appendix~C.
\end{IEEEproof}

Overall, we evaluate $T_{1}^{\textrm{NL}}$ as
\begin{eqnarray}
\hspace{-0.5cm} T_{1}^{\textrm{NL}}& = & \int_{0}^{y_{1}} \sum\limits_{k=0}^{\infty} \sum\limits_{m = 0}^{k} \ J(m,k) \ \gamma^{k-m}(-1)^{k-m} \frac{\partial^{k-m} [ \mathscr{L}_{I_{r}}(\frac{\gamma r^{\alpha^{{\rm {NL}}}}}{PA^{{\rm {NL}}}})}{\partial \gamma^{k-m}} f_{R,1}^{\textrm{NL}}\left(r\right) | 0<r\leq y_{1}]dr \nonumber \\
& & \hspace{-0.8cm} + \int_{y_{1}}^{d_{1}}\sum\limits_{k=0}^{\infty} \sum\limits_{m = 0}^{k} \ J(m,k) \ \gamma^{k-m}(-1)^{k-m} \frac{\partial^{k-m} [ \mathscr{L}_{I_{r}}(\frac{\gamma r^{\alpha^{{\rm {NL}}}}}{PA^{{\rm {NL}}}})}{\partial \gamma^{k-m}} f_{R,1}^{\textrm{NL}}\left(r\right) | y_{1}<r\leq d_{1}]dr,\label{eq:T_1_UAS1_NLoS_final}
\end{eqnarray}
where $f_{R,1}^{\textrm{NL}}\left(r\right)$ is computed using (\ref{eq:spec_geom_dis_PDF_UAS1_NLoS_seg1_case1}) and (\ref{eq:spec_geom_dis_PDF_UAS1_NLoS_seg1_case2}), 
and $\mathscr{L}_{I_{r}}\left(\frac{\gamma r^{\alpha^{{\rm {NL}}}}}{PA^{{\rm {NL}}}}\right)$ is given by (\ref{eq:Lemma_4-1}) and (\ref{eq:Lemma_4-2}).

\subsection{The Computation of $T_{2}^{\textrm{L}}$\label{sub:The-Computation-of-T2L}}

From Theorem~\ref{thm:p_cov_UAS1}, 
$T_{2}^{\textrm{L}}$ is computed as 
\begin{equation}
T_{2}^{\textrm{L}}=\int_{d_{1}}^{\infty}\sum\limits_{k=0}^{\infty} \sum\limits_{m = 0}^{k} \ J(m,k) \ \gamma^{k-m}(-1)^{k-m} \frac{\partial^{k-m} \mathscr{L}_{I_{r}}(\frac{\gamma}{P \zeta_{n}^{L}(r)})}{\partial \gamma^{k-m}} f_{R,2}^{\textrm{L}}\left(r\right)dr.\label{eq:T_2_UAS1_LoS}
\end{equation}

According to Theorem~\ref{thm:p_cov_UAS1} and (\ref{eq:LoS_Prob_func_linear}), 
the $f_{R,2}^{\textrm{L}}\left(r\right)$ can be written as
\begin{eqnarray}
f_{R,2}^{\textrm{L}}\left(r\right) & = & \exp\left(-\int_{0}^{r_{1}}\lambda\left(1-\textrm{Pr}^{\textrm{L}}\left(u\right)\right)2\pi udu\right)\exp\left(-\int_{0}^{r}\lambda\textrm{Pr}^{\textrm{L}}\left(u\right)2\pi udu\right)\times0\times2\pi r\lambda\nonumber \\
 & = & 0,\quad\left(r>d_{1}\right).\label{eq:spec_geom_dis_PDF_UAS1_LoS_seg2}
\end{eqnarray}

\subsection{The Computation of $T_{2}^{\textrm{NL}}$\label{sub:The-Computation-of-T2NL}}

From Theorem~\ref{thm:p_cov_UAS1}, 
$T_{2}^{\textrm{NL}}$ is computed as
\begin{eqnarray}
T_{2}^{\textrm{NL}} & = & \int_{d_{1}}^{\infty}\sum\limits_{k=0}^{\infty} \sum\limits_{m = 0}^{k} \ J(m,k) \ \gamma^{k-m}(-1)^{k-m} \frac{\partial^{k-m} \mathscr{L}_{I_{r}}(\frac{\gamma}{P \zeta_{n}^{NL}(r)})}{\partial \gamma^{k-m}} f_{R,2}^{\textrm{NL}}\left(r\right)dr\nonumber \\
 & \overset{(a)}{=} & \int_{d_{1}}^{\infty} \sum\limits_{k=0}^{\infty} \sum\limits_{m = 0}^{k} \ J(m,k) \ \gamma^{k-m}(-1)^{k-m} \frac{\partial^{k-m} \mathscr{L}_{I_{r}}(\frac{\gamma r^{\alpha^{{\rm {NL}}}}}{PA^{{\rm {NL}}}})}{\partial \gamma^{k-m}} f_{R,2}^{\textrm{NL}}\left(r\right)dr,\label{eq:T_2_UAS1_NLoS}
\end{eqnarray}
where $\zeta_{2}^{\textrm{NL}}\left(r\right)=A^{{\rm {NL}}}r^{-\alpha^{{\rm {NL}}}}$ from (\ref{eq:PL_BS2UE_2slopes}) is plugged into the step (a) of (\ref{eq:T_2_UAS1_NLoS}). 

Furthermore, based on Theorem~\ref{thm:p_cov_UAS1} and (\ref{eq:LoS_Prob_func_linear}), 
the $f_{R,2}^{\textrm{NL}}\left(r\right)$ can be written as
\begin{eqnarray}
f_{R,2}^{\textrm{NL}}\left(r\right)\hspace{-0.3cm} & = & \hspace{-0.3cm}\exp\left(-\int_{0}^{d_{1}}\lambda\left(1-\frac{u}{d_{1}}\right)2\pi udu\right)\exp\left(-\int_{0}^{d_{1}}\lambda\frac{u}{d_{1}}2\pi udu-\int_{d_{1}}^{r}\lambda2\pi udu\right)2\pi r\lambda\nonumber \\
\hspace{-0.3cm} & = & \hspace{-0.3cm}\exp\left(-\pi\lambda r^{2}\right)2\pi r\lambda,\quad\left(r>d_{1}\right).\label{eq:spec_geom_dis_PDF_UAS1_NLoS_seg2}
\end{eqnarray}

In the following, Lemma~\ref{lem:laplace_term_UAS1_NLoS_seg2} is proposed in order to to calculate $\mathscr{L}_{I_{r}}\left(\frac{\gamma r^{\alpha^{{\rm {NL}}}}}{PA^{{\rm {NL}}}}\right)$ in (\ref{eq:T_2_UAS1_NLoS}) for the range of $r>d_{1}$. 

\begin{lem}
\label{lem:laplace_term_UAS1_NLoS_seg2}$\mathscr{L}_{I_{r}}\left(\frac{\gamma r^{\alpha^{{\rm {NL}}}}}{PA^{{\rm {NL}}}}\right)$ in the range of $r>d_{1}$ can be computed as
\begin{eqnarray}
\mathscr{L}_{I_{r}}\left(\frac{\gamma r^{\alpha^{{\rm {NL}}}}}{PA^{{\rm {NL}}}}\right) & = & \nonumber \\
& & \hspace{-3.3cm}{\rm exp}\left(-2\pi\lambda \ \rho_{2}\left(\alpha^{NL},1,(1-K)\left(\gamma r^{\alpha^{\rm NL}}\right)^{-1},d_{1}\right)\right) \ \times {\rm exp} \biggl(-2\pi\lambda \left(\gamma r^{\alpha^{\rm NL}}\right)^{-1} \left(1-K-e^{-K}\right) \nonumber \\
& & \hspace{-3.3cm} \rho_{2}\left(\alpha^{NL},\alpha^{NL}+1,(1-K)\left(\gamma r^{\alpha^{\rm NL}}\right)^{-1},d_{1}\right)\biggr),\quad\left(r>d_{1}\right),\label{eq:Lemma_5}
\end{eqnarray}
where $\rho_{2}\left(\alpha,\beta,t,d\right)$ is defined in (\ref{eq:rou2_func}).
\end{lem}

\begin{IEEEproof}
See Appendix~D.
\end{IEEEproof}

Overall, we evaluate $T_{2}^{\textrm{NL}}$ as
\begin{equation}
T_{2}^{\textrm{NL}}=\int_{d_{1}}^{\infty} \sum\limits_{k=0}^{\infty} \sum\limits_{m = 0}^{k} \ J(m,k) \ \gamma^{k-m}(-1)^{k-m} \frac{\partial^{k-m} \mathscr{L}_{I_{r}}(\frac{\gamma r^{\alpha^{{\rm {NL}}}}}{PA^{{\rm {NL}}}})}{\partial \gamma^{k-m}} f_{R,2}^{\textrm{NL}}\left(r\right)dr.\label{eq:T_2_UAS1_NLoS_final}
\end{equation}
where $f_{R,2}^{\textrm{NL}}\left(r\right)$ and $\mathscr{L}_{I_{r}}\left(\frac{\gamma r^{\alpha^{{\rm {NL}}}}}{PA^{{\rm {NL}}}}\right)$ are computed by (\ref{eq:spec_geom_dis_PDF_UAS1_NLoS_seg2}) and (\ref{eq:Lemma_5}), respectively.

\subsection{The Results of $p^{\textrm{cov}}\left(\lambda,\gamma\right)$ and $A^{\textrm{ASE}}\left(\lambda,\gamma_{0}\right)$}

Based on the obtained derivations, 
the probability of coverage can be written as
\begin{equation}
p^{\textrm{cov}}\left(\lambda,\gamma\right)=T_{1}^{\textrm{L}}+T_{1}^{\textrm{NL}}+T_{2}^{\textrm{NL}},\label{eq:spec_p_cov_UAS1_final}
\end{equation}

Plugging $p^{\textrm{cov}}\left(\lambda,\gamma\right)$ into (\ref{eq:cond_SINR_PDF}),
the area spectral efficiency $A^{\textrm{ASE}}\left(\lambda,\gamma_{0}\right)$ can be obtained.

\section{Simulation and Discussion\label{sec:Simulation-and-Discussion}}

In this section, we use numerical results to study the performance of dense SCNs under the Rician fading channel,
and validate the accuracy of our analysis. 
Table~\ref{tab:simulation_setting} lists the simulation parameters.

\begin{table}[t]
\centering
\caption{Simulation Settings}{
\begin{tabular}{@{}lcc@{}}
\toprule
\bf{Parameter} & \bf{Values} \cite{TR36.828}\\
\midrule
$\alpha^{\rm {L}}$ & 2.09     \\
$\alpha^{\rm {NL}}$ & 3.75  \\
$A^{\rm {L}}$ & $10^{-10.38}$ \\
$A^{\rm {NL}}$ & $10^{-14.54}$ \\
$d_{1}$ & 0.3 km \\
P & 24 dBm \\
$N_{0}$ & -95 dBm \\
\toprule
\end{tabular}}
\label{tab:simulation_setting}
\end{table}

\subsection{Validation of the Analytical Results of $p^{\textrm{cov}}\left(\lambda,\gamma\right)$ for 3GPP Case\label{sub:Sim-p-cov-3GPP-Case}}

\begin{figure*}[t]
\centering
\includegraphics[scale=0.85]{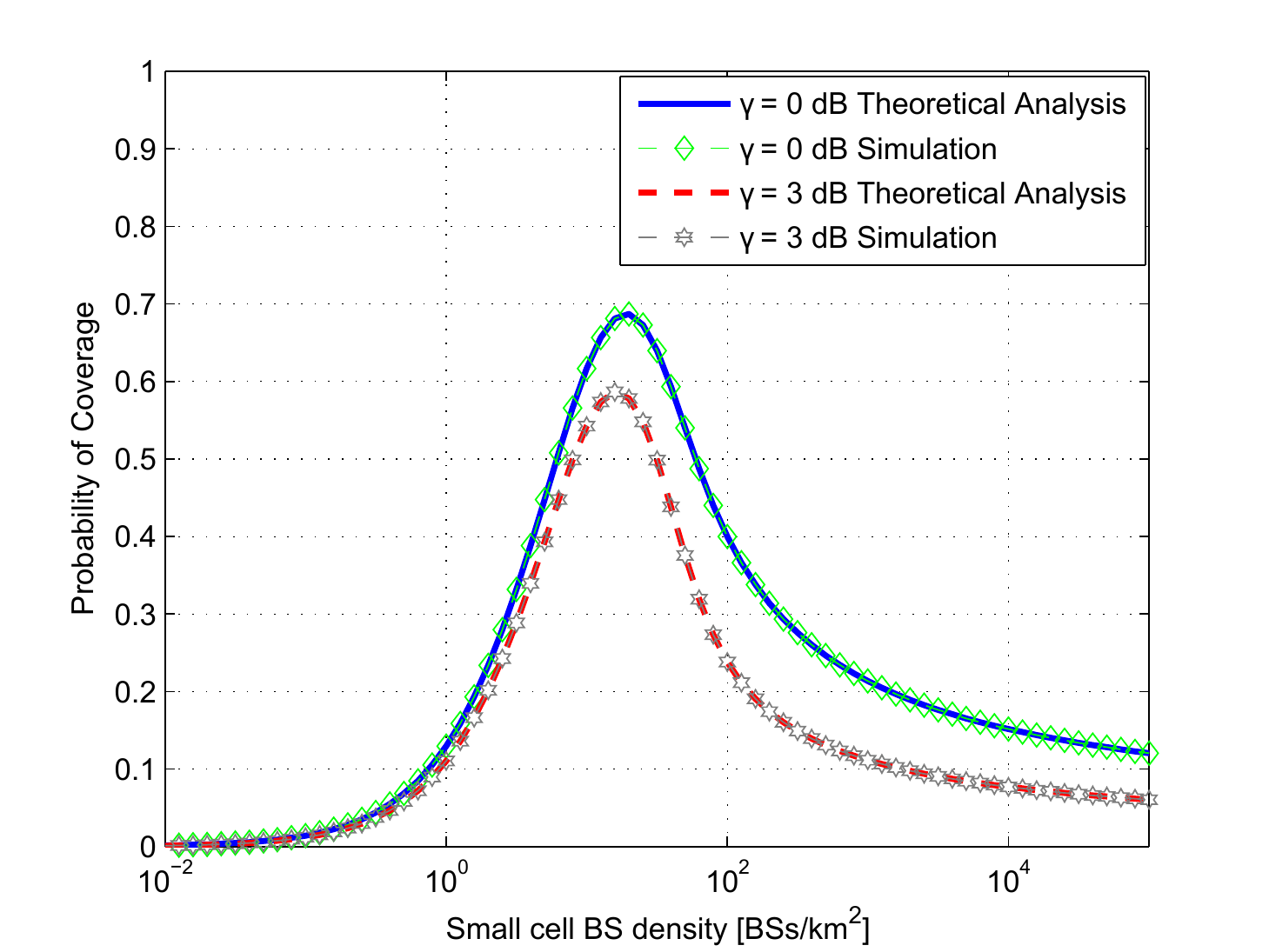}
\caption{The probability of coverage versus BS denisty.}
\label{fig:p_cov}
\vspace{-0.3cm}
\end{figure*}

Fig.~\ref{fig:p_cov} shows the results of $p^{\textrm{cov}}\left(\lambda,\gamma\right)$ for different SINR thresholds of $\gamma = 0 \ \rm dB$ and $\gamma = 3 \ \rm dB$. 
First, it is important to note that the theoretical analysis results match well with the simulation results, 
and hence we only show theoretical results in the sequel. 
Fig.~\ref{fig:p_cov} shows that the probability of coverage in the case of Ricean fading follows the same trend as in the case of Rayleigh fading presented in~\cite{LOS_NLOS_Trans}. 
More specifically, the probability of coverage initially increases as the BS density increases. 
However, once the BS density exceeds a certain threshold, 
i.e.,  $\lambda > \lambda_{1}$ $(e.g., \lambda_{1} = 100 \tfrac{\rm BSs}{\rm km^{2}} in Fig.~\ref{fig:p_cov})$, 
the probability of coverage starts to decline. 
This can be explained as follows. 
When the BS density is lower than $\lambda_{1}$, 
the network behaviour is noise limited and thus there is a rapid increase in coverage probability with the BS density. 
However, once the network becomes denser and the density of BSs is larger than $\lambda_{1}$, 
then a large number of interfering signals transit from NLOS to  LOS,
and hence the increase in interference power cannot be counterbalanced by the increase in signal power,
which was already LOS. 
Note that further densification beyond $\lambda_{1}$ results in slower decline rate in coverage probability, 
since both signals corresponding to interfering and serving BSs are LOS dominated. 

\begin{figure*}[t]
\centering
\includegraphics[scale=0.85]{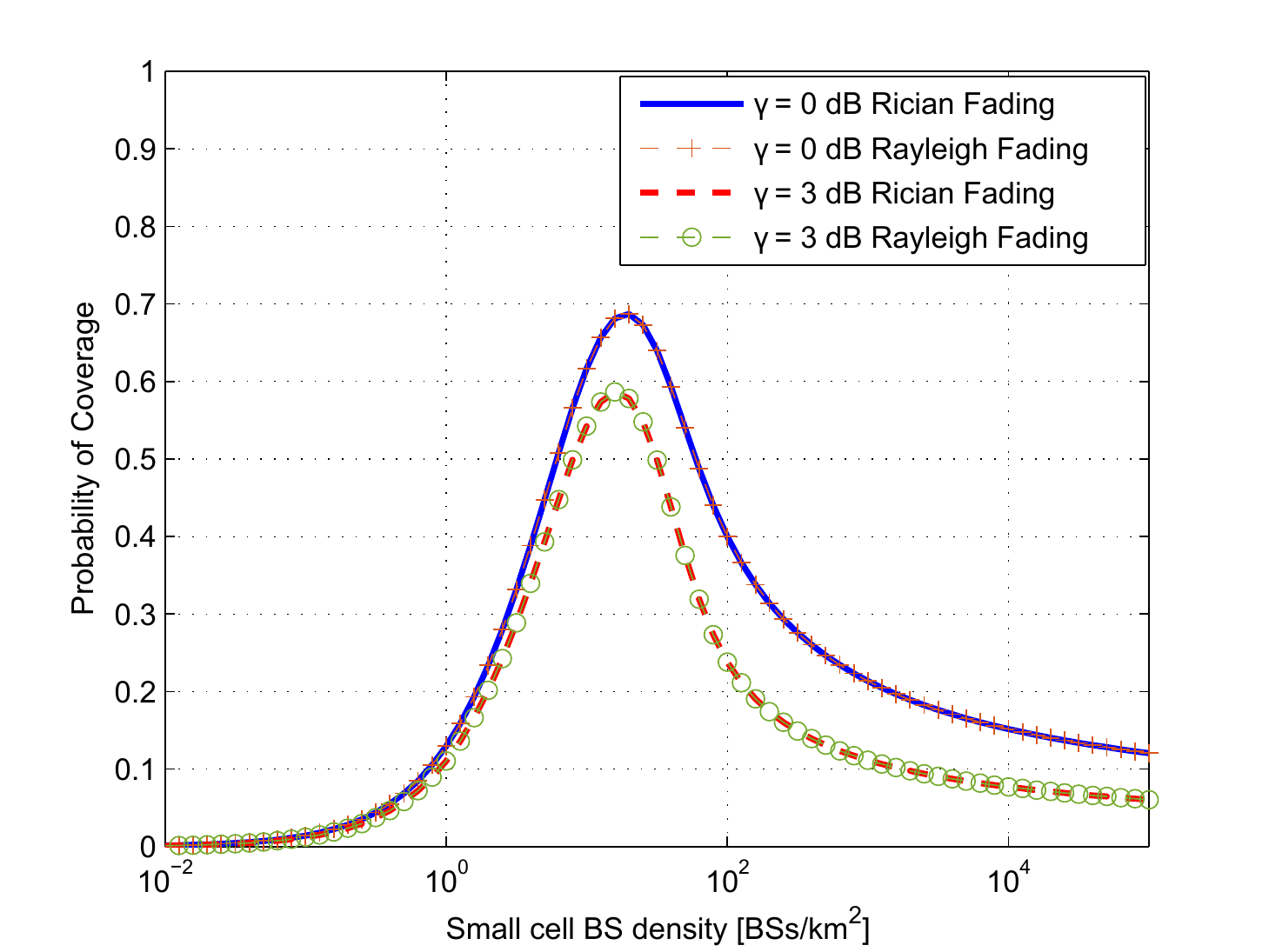}
\caption{Compariosn of the probability of coverage under Rayleigh and Rician fading.}
\label{fig:p_cov_comp}
\vspace{-0.3cm}
\end{figure*}

Comparing the probability of coverage results in this paper with Rician fading with those in~\cite{LOS_NLOS_Trans} with Rayleigh fading as shown in Fig. \ref{fig:p_cov_comp}, 
it can be concluded that the impact of Rician fading on the probability of coverage is negligible. 
The difference in coverage probability is less than 0.02 for all BS densities.
This is because
 the NLOS to LOS transition is in the order of 15-20 dB according to the 3GPP path loss functions \cite{TR36.828},
 while that of Rayleigh to Rician is in the order of $\sim$ 3 dB. 
Hence, Rayleigh or Rician fading makes little difference against this abrupt change of interference strength.

\subsection{Discussion on the Analytical Results of $A^{\textrm{ASE}}\left(\lambda,\gamma_{0}\right)$ for 3GPP Case\label{sub:Sim-ASE-3GPP-Case}}

\begin{figure*}[t]
\centering
\includegraphics[scale=0.85]{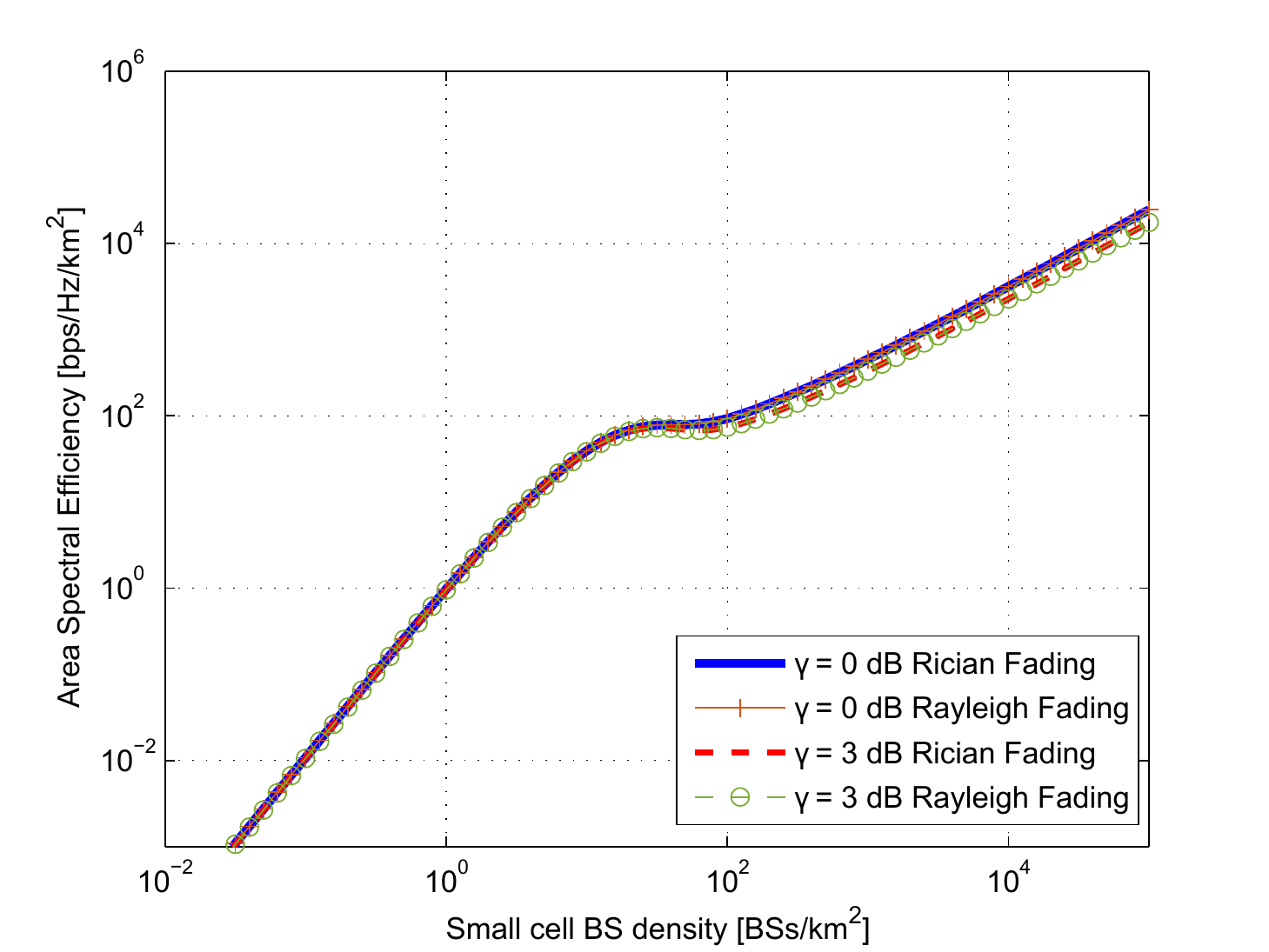}
\caption{The area spectral efficiency versus BS denisty.}
\label{fig:ASE}
\vspace{-0.3cm}
\end{figure*}

Fig.~\ref{fig:ASE} shows the ASE for different SINR thresholds of $\gamma = 0 \ \rm dB$ and $\gamma = 3 \ \rm dB$. 
Note that the ASE results are derived based on the results from the probability of coverage presented in (\ref{eq:ASE_def}). 
Similar to the observed trend for the probability of coverage, 
the ASE trend also shows three phases. 
In the first phase, 
when the BS density is lower than $\lambda_{1}$, 
the ASE increases with the BS density as coverage holes are mitigated. 
In the second phase, 
when the BS density exceeds $\lambda_{1}$, 
the ASE suffers from a slower growth pace or even a decrease 
due to the decline in probability of coverage originated by the transition of a large number of interfering signals transit from NLOS to  LOS. 
In the third phase, 
when all interfering signals has transited to LOS,
the ASE starts to linearly increase with BS density since the network has become statistically stable with all interfering and serving BSs being LOS dominated. 

Comparing the ASE results in this paper with Rician fading with those in~\cite{LOS_NLOS_Trans} with Rayleigh fading as shown in Fig. \ref{fig:ASE}, 
it can be concluded that the impact of Rician fading on the ASE is negligible with a peak Rician to Rayleigh gain of about 1.02x at a BS density of 15.85 $\frac{BSs}{km^{2}}$.
The reason of this conclusion has been explained before,
 i.e., the power variation of the NLOS to LOS transition is in the order of 15-20 dB according to the 3GPP path loss functions \cite{TR36.828}, 
while that of Rayleigh to Rician is in the order of $\sim$ 3 dB. 
Hence, Rayleigh or Rician fading makes little difference against this abrupt change of interference strength.

\section{Conclusion\label{sec:Conclusion}}

In this paper, we have considered the Rician fading channel in a path loss model that incorporates both LOS and NLOS transmissions in a dense SCN. 
Results show that Rician fading has a negligible impact compared to Rayleigh one on the system performance, 
indicating that the LOS and NLOS path loss characteristics and not the multi-path fading ones dominate the SCN performance in a single input single output scenario. 
Similar to the previous observations under the Rayleigh fading model, 
our results show that,
when the density of BSs exceeds a threshold, 
the ASE starts to suffer from a slow growth or even a decrease for a given BS density range. 
The intuition is as follows.
Network densification causes a transition from NLOS to LOS for a large number of interference signals
as well as a channel diversity loss as LOS dominates.
However, due to the dominance of the path loss characteristics over the multi-path fading ones in dense SCNs, 
the interference power increases faster than the signal power,
degrading the user SINR. 

\section*{Appendix~A: Proof of Theorem~\ref{thm:p_cov_UAS1} \label{sec:Appendix-A}}

To compute $p^{\textrm{cov}}\left(\lambda,\gamma\right)$, we first need to calculate the distance PDFs for the corresponding events of the typical UE being associated with a BS with either a LOS or NLOS path. Recalling from (\ref{eq:Coverage_Prob_def}) and (\ref{eq:SINR}), the $p^{\textrm{cov}}\left(\lambda,\gamma\right)$ can be computed as

\noindent 
\begin{eqnarray}
p^{\textrm{cov}}\left(\lambda,\gamma\right)\hspace{-0.3cm} & \stackrel{\left(a\right)}{=} & \hspace{-0.3cm}\int_{r>0}\textrm{Pr}\left[\left.\mathrm{SINR}>\gamma\right|r\right]f_{R}\left(r\right)dr\nonumber \\
\hspace{-0.3cm} & = & \hspace{-0.3cm}\int_{r>0}\textrm{Pr}\left[\frac{P\zeta\left(r\right)h}{I_{r}}>\gamma\right]f_{R}\left(r\right)dr\nonumber \\
\hspace{-0.3cm} & = & \hspace{-0.3cm}\int_{0}^{d_{1}}\textrm{Pr}\left[\frac{P\zeta_{1}^{\textrm{L}}\left(r\right)h}{I_{r}}>\gamma\right]\hspace{-0.1cm}f_{R,1}^{\textrm{L}}\left(r\right)dr\hspace{-0.1cm}+\hspace{-0.1cm}\int_{0}^{d_{1}}\textrm{Pr}\left[\frac{P\zeta_{1}^{\textrm{NL}}\left(r\right)h}{I_{r}}>\gamma\right]\hspace{-0.1cm}f_{R,1}^{\textrm{NL}}\left(r\right)dr\nonumber \\
\hspace{-0.3cm} &  & \hspace{-0.3cm}+\cdots\nonumber \\
\hspace{-0.3cm} &  & \hspace{-0.3cm}+\int_{d_{N-1}}^{\infty}\textrm{Pr}\left[\frac{P\zeta_{N}^{\textrm{L}}\left(r\right)h}{I_{r}}>\gamma\right]\hspace{-0.1cm}f_{R,N}^{\textrm{L}}\left(r\right)dr\hspace{-0.1cm}+\hspace{-0.1cm}\int_{d_{N-1}}^{\infty}\textrm{Pr}\left[\frac{P\zeta_{N}^{\textrm{NL}}\left(r\right)h}{I_{r}}>\gamma\right]\hspace{-0.1cm}f_{R,N}^{\textrm{NL}}\left(r\right)dr\nonumber \\
\hspace{-0.3cm} & \stackrel{\bigtriangleup}{=} & \hspace{-0.3cm}\sum_{n=1}^{N}\left(T_{n}^{\textrm{L}}+T_{n}^{\textrm{NL}}\right),\label{eq:p_cov_general_form}
\end{eqnarray}
where $f_{R,n}^{\textrm{L}}\left(r\right)$ and $f_{R,n}^{\textrm{NL}}\left(r\right)$ refer to the piece-wise densities of the RVs $R_{n}^{\textrm{L}}$ and $R_{n}^{\textrm{NL}}$, respectively. $R_{n}^{\textrm{L}}$ and $R_{n}^{\textrm{NL}}$ also denote the distances that the UE is associated with a BS with a LOS path and a NLOS path, respectively with the corresponding events are assumed to be disjoint. 

In the following, we present two events in order to calculate $f_{R,n}^{\textrm{L}}\left(r\right)$ in~(\ref{eq:p_cov_general_form}). 
\begin{itemize}
\item Event $B^{\textrm{L}}$: The nearest BS with a LOS path to the UE, is placed at distance $X^{\textrm{L}}$. According to~\cite{Jeff's work 2011}, the complementary cumulative distribution function (CCDF) of $X^{\textrm{L}}$ can be written as $\bar{F}_{X}^{\textrm{L}}\left(x\right)=\exp\left(-\int_{0}^{x}\textrm{Pr}^{\textrm{L}}\left(u\right)2\pi u\lambda du\right)$.

The PDF of $X^{\textrm{L}}$ can then be obtained by taking the derivative of $\left(1-\bar{F}_{X}^{\textrm{L}}\left(x\right)\right)$ with regard to $x$ as\\
\begin{equation}
f_{X}^{\textrm{L}}\left(x\right)=\exp\left(-\int_{0}^{x}\textrm{Pr}^{\textrm{L}}\left(u\right)2\pi u\lambda du\right)\textrm{Pr}^{\textrm{L}}\left(x\right)2\pi x\lambda.\label{eq:PDF_X_BL}
\end{equation}

\item Event $C^{\textrm{NL}}$ conditioned on the value of $X^{\textrm{L}}$: Given that $X^{\textrm{L}}=x$, the UE is associated with the nearest BS with a LOS path placed at distance $X^{\textrm{L}}$, giving the smallest path loss (i.e., the largest $\zeta\left(r\right)$) from such BS to the UE. To ensure that the UE is associated with
such LOS BS at distance $X^{\textrm{L}}=x$, there must no BS with a NLOS path inside the disk centered on the UE with a radius of $x_{1}<x$ to outperform such LOS BS at distance $X^{\textrm{L}}=x$, where $x_{1}$ satisfies $x_{1}=\underset{x_{1}}{\arg}\left\{ \zeta^{\textrm{NL}}\left(x_{1}\right)=\zeta^{\textrm{L}}\left(x\right)\right\} $. According to~\cite{Jeff's work 2011}, such conditional probability of $C^{\textrm{NL}}$ on condition of $X^{\textrm{L}}=x$ can be written as\\
\begin{equation}
\textrm{Pr}\left[\left.C^{\textrm{NL}}\right|X^{\textrm{L}}=x\right]=\exp\left(-\int_{0}^{x_{1}}\left(1-\textrm{Pr}^{\textrm{L}}\left(u\right)\right)2\pi u\lambda du\right).\label{eq:cond_prob_CNL}
\end{equation}

\end{itemize}
Note that Event $B^{\textrm{L}}$ guarantees that the path loss value $\zeta^{\textrm{L}}\left(x\right)$ associated with \emph{an arbitrary LOS BS} is always smaller than that associated with \emph{the considered LOS BS} at distance $X^{\textrm{L}}=x$. Moreover, conditioned on $X^{\textrm{L}}=x$, Event $C^{N\textrm{L}}$ guarantees that the path loss value $\zeta^{\textrm{NL}}\left(x\right)$ associated with \emph{an arbitrary NLOS BS} must always be smaller than that associated with \emph{the considered LOS BS} at distance $x$.

Another Event that has to be taken into account is the one that the UE is associated with a BS with a LOS path where the BS is placed at distance $R^{\textrm{L}}$. The CCDF of $R^{\textrm{L}}$, denoted by $\bar{F}_{R}^{\textrm{L}}\left(r\right)$, is derived as

\noindent 
\begin{eqnarray}
\bar{F}_{R}^{\textrm{L}}\left(r\right) & = & \textrm{Pr}\left[R^{\textrm{L}}>r\right]\nonumber \\
 & \stackrel{\left(a\right)}{=} & \textrm{E}_{\left[X^{\textrm{L}}\right]}\left\{ \textrm{Pr}\left[\left.R^{\textrm{L}}>r\right|X^{\textrm{L}}\right]\right\} \nonumber \\
 & = & \int_{0}^{+\infty}\textrm{Pr}\left[\left.R^{\textrm{L}}>r\right|X^{\textrm{L}}=x\right]f_{X}^{\textrm{L}}\left(x\right)dx\nonumber \\
 & \stackrel{\left(b\right)}{=} & \int_{0}^{r}0\times f_{X}^{\textrm{L}}\left(x\right)dx+\int_{r}^{+\infty}\textrm{Pr}\left[\left.C^{\textrm{NL}}\right|X^{\textrm{L}}=x\right]f_{X}^{\textrm{L}}\left(x\right)dx\nonumber \\
 & = & \int_{r}^{+\infty}\textrm{Pr}\left[\left.C^{\textrm{NL}}\right|X^{\textrm{L}}=x\right]f_{X}^{\textrm{L}}\left(x\right)dx,\label{eq:CCDF_SINR_req_UAS1_LoS}
\end{eqnarray}

\noindent where $\mathbb{E}_{\left[X\right]}\left\{ \cdot\right\} $ in the step (a) of (\ref{eq:CCDF_SINR_req_UAS1_LoS}) represent the expectation operation taking the expectation over the variable $X$
and the step (b) of (\ref{eq:CCDF_SINR_req_UAS1_LoS}) is valid since $\textrm{Pr}\left[\left.R^{\textrm{L}}>r\right|X^{\textrm{L}}=x\right]=0$ when $0<x\leq r$ and the conditional event $\left[\left.R^{\textrm{L}}>r\right|X^{\textrm{L}}=x\right]$ is equivalent to the conditional event $\left[\left.C^{\textrm{NL}}\right|X^{\textrm{L}}=x\right]$ when $x>r$. In order to obtain the PDF of $R^{\textrm{L}}$, we can take the derivative of $\left(1-\bar{F}_{R}^{\textrm{L}}\left(r\right)\right)$ with regard to $r$ which results in

\noindent 
\begin{equation}
f_{R}^{\textrm{L}}\left(r\right)=\textrm{Pr}\left[\left.C^{\textrm{NL}}\right|X^{\textrm{L}}=r\right]f_{X}^{\textrm{L}}\left(r\right).\label{eq:geom_dis_PDF_total_UAS1_LoS}
\end{equation}

\noindent Considering the distance range of $\left(d_{n-1}<r\leq d_{n}\right)$, the segment of $f_{R,n}^{\textrm{L}}\left(r\right)$ from $f_{R}^{\textrm{L}}\left(r\right)$ can be derived as 

\noindent 
\begin{eqnarray}
f_{R,n}^{\textrm{L}}\left(r\right) & = & \exp\left(-\int_{0}^{r_{1}}\left(1-\textrm{Pr}^{\textrm{L}}\left(u\right)\right)2\pi u\lambda du\right)\nonumber \\
 &  & \times\exp\left(-\int_{0}^{r}\textrm{Pr}^{\textrm{L}}\left(u\right)2\pi u\lambda du\right)\textrm{Pr}_{n}^{\textrm{L}}\left(r\right)2\pi r\lambda,\quad\left(d_{n-1}<r\leq d_{n}\right),\label{eq:geom_dis_PDF_UAS1_LoS}
\end{eqnarray}

\noindent where $r_{1}=\underset{r_{1}}{\arg}\left\{ \zeta^{\textrm{NL}}\left(r_{1}\right)=\zeta_{n}^{\textrm{L}}\left(r\right)\right\} $. 

Having obtained $f_{R,n}^{\textrm{L}}\left(r\right)$, we move on to evaluate $\textrm{Pr}\left[\frac{P\zeta_{n}^{\textrm{L}}\left(r\right)h}{I_{r}}>\gamma\right]$ in~(\ref{eq:p_cov_general_form}) as

\noindent 
\vspace{0.05cm}
\begin{equation}
{\rm {Pr}}[\frac{P \zeta_{n}^{L}(r) h}{I_{r}} > \gamma] = 1 - {\rm {Pr}}[\frac{P \zeta_{n}^{L}(r) h}{I_{r}} < \gamma]
\label{eq:cov_prob}
\end{equation}
where  ${\rm {Pr}}[\frac{P \zeta_{n}^{L}(r) h}{I_{r}} > \gamma]$ and ${\rm {Pr}}[\frac{P \zeta_{n}^{L}(r) h}{I_{r}} < \gamma]$ refer to the CCDF and CDF of SINR, respectively. Worth reminding that for tractability of analysis, we have considered an interference limited scenario.

The interference is normalized with respect to $P \zeta_{n}^{L}(r)$ and therefore the normalized interference is defined as $I_{rn} = \frac{I_{r}}{P \zeta_{n}^{L}(r)}$. Hence, (\ref{eq:cov_prob}) can be expressed as

\begin{equation}
{\rm {Pr}}[\frac{h} {I_{rn}} > \gamma] = 1 - {\rm {Pr}}[\frac {h} {I_{rn}} < \gamma]
\label{eq:int_norm}
\end{equation}

Subsequently, the coverage probability can be computed as
\vspace{-0.2cm}

\begin{equation}
{\rm {Pr}}[\frac{h}{I_{rn}} > \gamma] = 1 - \int \int_{\frac{x}{y}< \gamma} f_{h}(x) f_{I_{rn}}(y) \ dx \ dy = 1 - \int_{0}^{\infty} F_{h}(\gamma y) f_{I_{rn}}(y) \ dy
\label{eq:prob_cov}
\end{equation}
where $f_{h}(x)$ and $F_{h}(x)$ denote the PDF and CDF of random variable $h$, respectively. Assuming that the random variable $h$ is Rician distributed, its PDF is given by

\begin{equation}
f_{h}(x) = \frac{(K+1)e^{-K}}{\bar{x}} \ {\rm{exp}}(- \frac{(K+1)x}{\bar{x}}) \ I_{0}(\sqrt{\frac{4K(K+1)x}{\bar{x}}})
\end{equation}
where $K$ refers to the Rician $K$ factor, $I_{0}$ is the zeroth order first kind modified Bessel function and $\bar{x}$ refers to the expectation of $h$. Applying the series expansion from~\cite{Book_Integrals}, the $f_{h}(x)$ can be expressed as

\begin{equation}
f_{h}(x) = {\rm {exp}}(-K - x) \sum\limits_{k=0}^{\infty} \frac{(Kx)^{k}}{(k!)^{2}}
\end{equation}
and therefore, the CDF of $h$ can be derived from its PDF as

\begin{eqnarray}
\hspace{2cm} F_{h}(x) & = & e^{-K}\sum\limits_{k=0}^{\infty} \frac{K^{k}}{(k!)^{2}} \ \left(e^{-x} \sum\limits_{m=0}^{k}(-1)^{2m+1} \ m!  \ {k \choose m} x^{k-m} + k!\right) \nonumber \\
 & = & - \sum\limits_{k=0}^{\infty}\sum\limits_{m=0}^{k} \ J(m,k) \ x^{k-m} e^{-x} \ + \sum\limits_{k=0}^{\infty} \frac{K^{k}}{k!} e^{-K} \nonumber \\
 & = & - \sum\limits_{k = 0}^{\infty} \sum\limits_{m = 0}^{k} \ J(m,k) \ x^{k-m} e^{-x} + 1
\label{eq:CDF}
\end{eqnarray}
where $J(m,k) = \frac{e^{-K} K^{k} m! {k \choose m}}{(k!)^{2}}$ and $\sum\limits_{k=0}^{\infty}\frac{K^{k}}{k!}= e^{K}$ based on the combination of Taylor series.

By replacing (\ref{eq:CDF}) in (\ref{eq:prob_cov}), the coverage probability can be derived as

\begin{eqnarray}
{\rm {Pr}}[\frac{h}{I_{rn}}>\gamma] = \sum\limits_{k=0}^{\infty}\sum\limits_{m=0}^{k} \ J(m,k) \int_{0}^{\infty} (y\gamma)^{k-m} e^{-y\gamma} f_{I_{rn}}(y) \ dy\} \nonumber \\
 & \hspace{-11cm} = \sum\limits_{k=0}^{\infty} \sum\limits_{m=0}^{k} \ J(m,k) \ \gamma^{k-m} Q(\gamma,k-m)
\end{eqnarray}
where $Q(\tau,n) = \int_{0}^{\infty} y^{n} e^{-y\tau} f_{I_{rn}}(y) dy = (-1)^{n} \frac{\partial^{n} \mathscr{L}_{I_{rn}}(\tau)}{\partial \tau^{n}}$ and $n = 0,1,..,\infty$ \cite{Rician_Quek} \cite{Rician_Peng}. Also, note that $\int_{0}^{\infty} f_{I_{rn}}(y) \ dy = 1$. Fianlly, the coverage probability can be presented as

\begin{eqnarray}
{\rm {Pr}}[\frac{h}{I_{rn}}>\gamma] = \sum\limits_{k=0}^{\infty} \sum\limits_{m = 0}^{k} \ J(m,k) \ \gamma^{k-m}(-1)^{k-m} \ \frac{\partial^{k-m} \mathscr{L}_{I_{rn}}(\gamma)}{\partial \gamma^{k-m}}
\label{eq:Intf1}
\end{eqnarray}

Plugging $I_{r} = I_{rn} \ P \zeta_{n}^{L}(r)$ into (\ref{eq:Intf1}), we can derive 

\begin{eqnarray}
{\rm {Pr}}[\frac{P \zeta_{n}^{L}(r)h}{I_{r}}>\gamma] = \sum\limits_{k=0}^{\infty} \sum\limits_{m = 0}^{k} \ J(m,k) \ \gamma^{k-m}(-1)^{k-m} \ \frac{\partial^{k-m} \mathscr{L}_{I_{r}}(\frac{\gamma}{P \zeta_{n}^{L}(r)})}{\partial \gamma^{k-m}}
\label{eq:Intf2}
\end{eqnarray}

\noindent where $\mathscr{L}_{I_{r}}\left(s\right)$ is the Laplace transform of RV $I_{r}$ evaluated at $s$. 

Similarly, $f_{R,n}^{\textrm{NL}}\left(r\right)$ can also be computed. In this regard, we define the following two events.

\begin{itemize}
\item Event $B^{\textrm{NL}}$: The nearest BS with a NLOS path to the UE, is placed at distance $X^{\textrm{NL}}$. Similar to (\ref{eq:PDF_X_BL}), the PDF of $X^{\textrm{NL}}$ is given by\\
\begin{equation}
f_{X}^{\textrm{NL}}\left(x\right)=\exp\left(-\int_{0}^{x}\left(1-\textrm{Pr}^{\textrm{L}}\left(u\right)\right)2\pi u\lambda du\right)\left(1-\textrm{Pr}^{\textrm{L}}\left(x\right)\right)2\pi x\lambda.\label{eq:PDF_X_BNL}
\end{equation}

\item Event $C^{\textrm{L}}$ conditioned on the value of $X^{\textrm{NL}}$: Given that $X^{\textrm{NL}}=x$, the UE is associated with the nearest BS with a NLOS path palced at distance $X^{\textrm{NL}}$, which gives the smallest path loss (i.e., the largest $\zeta\left(r\right)$) from such BS to the UE. Consequently, there should be no BS with an
LOS path inside the disk centred on the UE with a radius of $x_{2}<x$, where $x_{2}$ satisfies $x_{2}=\underset{x_{2}}{\arg}\left\{ \zeta^{\textrm{L}}\left(x_{2}\right)=\zeta^{\textrm{NL}}\left(x\right)\right\} $. Similar to (\ref{eq:cond_prob_CNL}), such conditional probability of $C^{\textrm{L}}$ on condition of $X^{\textrm{NL}}=x$ can be expressed
as\\
\begin{equation}
\textrm{Pr}\left[\left.C^{\textrm{L}}\right|X^{\textrm{NL}}=x\right]=\exp\left(-\int_{0}^{x_{2}}\textrm{Pr}^{\textrm{L}}\left(u\right)2\pi u\lambda du\right).\label{eq:cond_prob_CL}
\end{equation}

\end{itemize}
Another Event that must be taken into account is the one that the UE is associated with a BS with a NLOS path and such BS is placed at distance $R^{\textrm{NL}}$. Similar to (\ref{eq:CCDF_SINR_req_UAS1_LoS}), the CCDF of $R^{\textrm{NL}}$, denoted by $\bar{F}_{R}^{\textrm{NL}}\left(r\right)$, can be computed as

\noindent 
\begin{eqnarray}
\bar{F}_{R}^{\textrm{NL}}\left(r\right) & = & \textrm{Pr}\left[R^{\textrm{NL}}>r\right]\nonumber \\
 & = & \int_{r}^{+\infty}\textrm{Pr}\left[\left.C^{\textrm{L}}\right|X^{\textrm{NL}}=x\right]f_{X}^{\textrm{NL}}\left(x\right)dx.\label{eq:CCDF_SINR_req_UAS1_NLoS}
\end{eqnarray}

\noindent The PDF of $R^{\textrm{NL}}$ can be obtained by taking the derivative of $\left(1-\bar{F}_{R}^{\textrm{NL}}\left(r\right)\right)$ with regard to $r$ which results in

\noindent 
\begin{equation}
f_{R}^{\textrm{NL}}\left(r\right)=\textrm{Pr}\left[\left.C^{\textrm{L}}\right|X^{\textrm{NL}}=r\right]f_{X}^{\textrm{NL}}\left(x\right).\label{eq:geom_dis_PDF_total_UAS1_NLoS}
\end{equation}

\noindent Considering the distance range of $\left(d_{n-1}<r\leq d_{n}\right)$, the segment of $f_{R,n}^{\textrm{NL}}\left(r\right)$ from $f_{R}^{\textrm{NL}}\left(r\right)$ can be derived as 

\noindent 
\begin{eqnarray}
\hspace{0.9cm} f_{R,n}^{\textrm{NL}}\left(r\right) & = & \exp\left(-\int_{0}^{r_{2}}\textrm{Pr}^{\textrm{L}}\left(u\right)2\pi u\lambda du\right)\nonumber \\
 &  & \times\exp\left(-\int_{0}^{r}\left(1-\textrm{Pr}^{\textrm{L}}\left(u\right)\right)2\pi u\lambda du\right)\left(1-\textrm{Pr}_{n}^{\textrm{L}}\left(r\right)\right)2\pi r\lambda,\left(d_{n-1}<r\leq d_{n}\right) \nonumber \\
 & &
\label{eq:geom_dis_PDF_UAS1_NLoS}
\end{eqnarray}

\noindent where $r_{2}=\underset{r_{2}}{\arg}\left\{ \zeta^{\textrm{L}}\left(r_{2}\right)=\zeta_{n}^{\textrm{NL}}\left(r\right)\right\} $. 

Similarly, $\textrm{Pr}\left[\frac{P\zeta_{n}^{\textrm{NL}}\left(r\right)h}{I_{r}}>\gamma\right]$ can be calculated as

\noindent 
\begin{eqnarray}
\textrm{Pr}\left[\frac{P\zeta_{n}^{\textrm{NL}}\left(r\right)h}{I_{r}}>\gamma\right] = \sum\limits_{k=0}^{\infty} \sum\limits_{m = 0}^{k} \ J(m,k) \ \gamma^{k-m}(-1)^{k-m} \frac{\partial^{k-m} \mathscr{L}_{I_{r}}(\frac{\gamma}{P \zeta_{n}^{NL}(r)})}{\partial \gamma^{k-m}}.\label{eq:Pr_SINR_req_wLT_UAS1_NLoS}
\end{eqnarray}

\section*{Appendix~B: Proof of Lemma~\ref{lem:laplace_term_UAS1_LoS_seg1}\label{sec:Appendix-B}}

In the following, we derive $\mathscr{L}_{I_{r}}\left(s\right)$ in the range of $0<r\leq d_{1}$ as

\noindent 
\begin{eqnarray}
\mathscr{L}_{I_{r}}\left(s\right) & = & \mathbb{E}_{\left[I_{r}\right]}\left\{ \left.\exp\left(-sI_{r}\right)\right|0<r\leq d_{1}\right\} \nonumber \\
 & = & \mathbb{E}_{\left[\Phi,\left\{ \beta_{i}\right\} ,\left\{ g_{i}\right\} \right]}\left\{ \left.\exp\left(-s\sum_{i\in\Phi/b_{o}}P\beta_{i}g_{i}\right)\right|0<r\leq d_{1}\right\} \nonumber \\
 & \overset{(a)}{=} & \exp\left(\left.-2\pi\lambda\int_{r}^{\infty}\left(1-\mathbb{E}_{\left[g\right]}\left\{ \exp\left(-sP\beta\left(u\right)g\right)\right\} \right)udu\right|0<r\leq d_{1}\right),\label{eq:laplace_term_LoS_UAS1_seg1_proof_eq1}
\end{eqnarray}

\noindent where the step (a) of (\ref{eq:laplace_term_LoS_UAS1_seg1_proof_eq1}) is obtained from~\cite{Jeff's work 2011}. 
 
Considering that $0<r \leq d_{1}$, ${\rm {E}}_{[g]}\{{\rm {exp}}(-sP\beta (u) g)\}$ in (\ref{eq:laplace_term_LoS_UAS1_seg1_proof_eq1}), must take into account the interference from both the LOS and NLOS paths. Note that the random varibale $g$ follows Rician distribution. Therefore, $\mathscr{L}_{I_{r}}(s)$ can be expressed as

\begin{eqnarray}
\mathscr{L}_{I_{r}}(s) & = & {\rm {exp}} \left(-2 \pi \lambda \int_{r}^{d_{1}} \left(1 - \frac{u}{d_{1}}\right) [1 - {\rm {E}}_{[g]} { {\rm {exp}}\left(-sPA^{L} u^{-\alpha^{L}} g\right)}] \ udu \right) \nonumber \\
& & \times {\rm {exp}} \biggl(-2 \pi \lambda \int_{r_{1}}^{d_{1}} \frac{u}{d_{1}} [1 - {\rm {E}}_{[g]} { {\rm {exp}} (-sP A^{NL} u^{-\alpha^{NL}} g)}] \ udu\biggr) \nonumber \\ 
& & \times {\rm {exp}} \biggl(-2\pi \lambda \int_{d_{1}}^{\infty} [1 - {\rm {E}}_{[g]} { {\rm {exp}} (-sPA^{NL} u^{-\alpha^{NL}} g)}] udu\biggr)
\label{eq:laplace2}
\end{eqnarray}

For sake of presentation, $sPA^{L}u^{-\alpha^{L}}$ is denoted by $M$ and hence ${\rm {E}}_{[g]} \{{\rm {exp}} (-Mg) \}$ is computed as
\begin{eqnarray}
{\rm {E}}_{[g]} { {\rm {exp}} (-Mg)} = \int_{0}^{\infty} {\rm {exp}}(-Mg) \ {\rm {exp}}(-K-g) \ \sum\limits_{k=0}^{\infty} \frac{(Kg)^{k}}{(k!)^{2}} \ dg
\label{eq:expec1}
\end{eqnarray}
where ${\rm {exp}}(-K-g) \sum\limits_{k=0}^{\infty} \frac{(Kg)^{k}}{(k!)^{2}}$ denotes the PDF of random variable $g$. According to Taylor series, it is realized that $\sum\limits_{k=0}^{\infty} \frac{K^k}{k!} = e^{K}$ and hence, (\ref{eq:expec1}) can be written as

\vspace{-1cm}
\begin{eqnarray}
{\rm {E}}_{[g]}\{ {\rm {exp}} (-Mg)\} & = & \int_{0}^{\infty} {\rm {exp}}(-Mg) {\rm {exp}}(-K-g) {\rm {exp}}(Kg) \ dg \nonumber \\
& = & {\rm {exp}}(-K) \int_{0}^{\infty} {\rm {exp}} (-g(1+M-K)) \ dg = \frac{{\rm {exp}} (-K)}{1+M-K}
\label{eq:expec_form}
\end{eqnarray}
Plugging $M = sPA^{L}u^{-\alpha^{L}}$ into (\ref{eq:expec_form}), the term $1 - {\rm {E}}_{[g]} {{\rm {exp}}(-sPA^{L}u^{-\alpha^{L}} g)}$ is derived as

\begin{eqnarray}
1 - {\rm {E}}_{[g]} {{\rm {exp}}(-sPA^{L}u^{-\alpha^{L}} g)} = \frac{1 + (sPA^{L})^{-1} u^{\alpha^{L}} - K(sPA^{L})^{-1} u^{\alpha^{L}} - (e^{K}sPA^{L})^{-1} u^{\alpha^{L}}}{1 + (sPA^{L})^{-1} u^{\alpha^{L}} - K(sPA^{L})^{-1} u^{\alpha^{L}}}
\end{eqnarray}
 
Similarly, the term $1 - {\rm {E}}_{[g]} \{{\rm {exp}}(-sPA^{NL}u^{-\alpha^{NL}} g)\}$ is computed and therefore, (\ref{eq:laplace2}) is written as

\begin{eqnarray}
\mathscr{L}_{I_{r}}(s) & = & \nonumber \\
& & \hspace{-2cm} {\rm {exp}}\Biggl(-2\pi \lambda \int_{r}^{d_{1}} \left(1-\frac{u}{d_{1}}\right) \left(\frac{1 + (sPA^{L})^{-1} u^{\alpha^{L}} - K(sPA^{L})^{-1} u^{\alpha^{L}} - (e^{K}sPA^{L})^{-1} u^{\alpha^{L}}}{1 + (sPA^{L})^{-1} u^{\alpha^{L}} - K(sPA^{L})^{-1} u^{\alpha^{L}}}\Biggr) udu\right) \nonumber
\\ & & \hspace{-2cm} \times \ {\rm {exp}}\Biggl(-2\pi \lambda \int_{r_{1}}^{d_{1}}\frac{u}{d_{1}} \left(\frac{1 + (sPA^{NL})^{-1} u^{\alpha^{NL}} - K(sPA^{NL})^{-1} u^{\alpha^{NL}} - (e^{K}sPA^{NL})^{-1} u^{\alpha^{NL}}}{1 + (sPA^{NL})^{-1} u^{\alpha^{NL}} - K(sPA^{NL})^{-1} u^{\alpha^{NL}}}\right) udu\Biggr) \nonumber
\\ & & \hspace{-2cm} \times \ {\rm {exp}}\Biggl(-2\pi \lambda \int_{d_{1}}^{\infty} \left(\frac{1 + (sPA^{NL})^{-1} u^{\alpha^{NL}} - K(sPA^{NL})^{-1} u^{\alpha^{NL}} - (e^{K}sPA^{NL})^{-1} u^{\alpha^{NL}}}{1 + (sPA^{NL})^{-1} u^{\alpha^{NL}} - K(sPA^{NL})^{-1} u^{\alpha^{NL}}}\right) udu\Biggr) \nonumber \\
& &
\label{eq:laplace_term_LoS_UAS1_seg1_proof_eq2}
\end{eqnarray}

Plugging $s=\frac{\gamma r^{\alpha^{\textrm{L}}}}{PA^{\textrm{L}}}$ into (\ref{eq:laplace_term_LoS_UAS1_seg1_proof_eq2}), and considering the definition of $\rho_{1}\left(\alpha,\beta,t,d\right)$ and $\rho_{2}\left(\alpha,\beta,t,d\right)$ in (\ref{eq:rou1_func}) and (\ref{eq:rou2_func}), we can obtain
$\mathscr{L}_{I_{r}}\left(\frac{\gamma r^{\alpha^{{\rm {L}}}}}{PA^{{\rm {L}}}}\right)$ as shown in (\ref{eq:Lemma_3}).

\section*{\noindent Appendix~C: Proof of Lemma~\ref{lem:laplace_term_UAS1_NLoS_seg1}\label{sec:Appendix-C}}

Similar to Appendix~B, we derive $\mathscr{L}_{I_{r}}\left(\frac{\gamma r^{\alpha^{{\rm {NL}}}}}{PA^{{\rm {NL}}}}\right)$ in the range of $0<r\leq y_{1}$ as 
\begin{eqnarray}
\mathcal{L}_{I_{r}}(\frac{\gamma r^{\alpha^{\rm NL}}}{PA^{\rm NL}}) = \nonumber \\
& \hspace{-2.8cm} {\rm {exp}}\Biggl(-2\pi \lambda \int_{r_{2}}^{d_{1}} (1-\frac{u}{d_{1}})
\frac{1 + (\frac{\gamma r^{\alpha^{\rm NL}}}{PA^{\rm NL}}
PA^{L})^{-1} u^{\alpha^{L}} - K(\frac{\gamma r^{\alpha^{\rm NL}}}{PA^{\rm NL}}
PA^{L})^{-1} u^{\alpha^{L}} - (e^{K}\frac{\gamma r^{\alpha^{\rm NL}}}{PA^{\rm NL}}
PA^{L})^{-1} u^{\alpha^{L}}}{1 + (\frac{\gamma r^{\alpha^{\rm NL}}}{PA^{\rm NL}}
PA^{L})^{-1} u^{\alpha^{L}} - K(\frac{\gamma r^{\alpha^{\rm NL}}}{PA^{\rm NL}}
PA^{L})^{-1} u^{\alpha^{L}}})u du\Biggr) \nonumber
\\ & \hspace{-3cm} \times \ {\rm {exp}}\Biggl(-2\pi \lambda \int_{r}^{d_{1}}\frac{u}{d_{1}} 
 \frac{1 + (\frac{\gamma r^{\alpha^{\rm NL}}}{PA^{\rm NL}}
PA^{NL})^{-1} u^{\alpha^{NL}} - K(\frac{\gamma r^{\alpha^{\rm NL}}}{PA^{\rm NL}}
PA^{NL})^{-1} u^{\alpha^{NL}} - (e^{K}\frac{\gamma r^{\alpha^{\rm NL}}}{PA^{\rm NL}}
PA^{NL})^{-1} u^{\alpha^{NL}}}{1 + (\frac{\gamma r^{\alpha^{\rm NL}}}{PA^{\rm NL}}
PA^{NL})^{-1} u^{\alpha^{NL}} - K(\frac{\gamma r^{\alpha^{\rm NL}}}{PA^{\rm NL}}
PA^{NL})^{-1} u^{\alpha^{NL}}} udu \Biggr) \nonumber 
\\ & \hspace{-3.2cm} \times \ {\rm {exp}} \Biggl(-2\pi \lambda \int_{d_{1}}^{\infty} \nonumber 
\frac{1 + (\frac{\gamma r^{\alpha^{\rm NL}}}{PA^{\rm NL}}
PA^{NL})^{-1} u^{\alpha^{NL}} - K(\frac{\gamma r^{\alpha^{\rm NL}}}{PA^{\rm NL}}
PA^{NL})^{-1} u^{\alpha^{NL}} - (e^{K}\frac{\gamma r^{\alpha^{\rm NL}}}{PA^{\rm NL}}
PA^{NL})^{-1} u^{\alpha^{NL}}}{1 + (\frac{\gamma r^{\alpha^{\rm NL}}}{PA^{\rm NL}}
PA^{NL})^{-1} u^{\alpha^{NL}} - K(\frac{\gamma r^{\alpha^{\rm NL}}}{PA^{\rm NL}}
PA^{NL})^{-1} u^{\alpha^{NL}}} udu \Biggr) \nonumber \\
& &.
\label{eq:proof_Lemma4_eq1}
\end{eqnarray}

Similarly, $\mathscr{L}_{I_{r}}\left(\frac{\gamma r^{\alpha^{{\rm {NL}}}}}{PA^{{\rm {NL}}}}\right)$
in the range of $y_{1}<r\leq d_{1}$ can be calculated by

\noindent 
\begin{eqnarray}
\mathscr{L}_{I_{r}}\left(\frac{\gamma r^{\alpha^{{\rm {NL}}}}}{PA^{{\rm {NL}}}}\right)= \nonumber \\
& \hspace{-3.5cm} {\rm {exp}}\Biggl(-2\pi \lambda \int_{r}^{d_{1}}\frac{u}{d_{1}}
\frac{1 + (\frac{\gamma r^{\alpha^{\rm NL}}}{PA^{\rm NL}}
PA^{NL})^{-1} u^{\alpha^{NL}} - K(\frac{\gamma r^{\alpha^{\rm NL}}}{PA^{\rm NL}}
PA^{NL})^{-1} u^{\alpha^{NL}} - (e^{K}\frac{\gamma r^{\alpha^{\rm NL}}}{PA^{\rm NL}}
PA^{NL})^{-1} u^{\alpha^{NL}}}{1 + (\frac{\gamma r^{\alpha^{\rm NL}}}{PA^{\rm NL}}
PA^{NL})^{-1} u^{\alpha^{NL}} - K(\frac{\gamma r^{\alpha^{\rm NL}}}{PA^{\rm NL}}
PA^{NL})^{-1} u^{\alpha^{NL}}} u du\Biggr) \nonumber \\
& \hspace{-3.5cm} \times \ {\rm {exp}}\Biggl(-2\pi \lambda \int_{d_{1}}^{\infty} 
\frac{1 + (\frac{\gamma r^{\alpha^{\rm NL}}}{PA^{\rm NL}}
PA^{NL})^{-1} u^{\alpha^{NL}} - K(\frac{\gamma r^{\alpha^{\rm NL}}}{PA^{\rm NL}}
PA^{NL})^{-1} u^{\alpha^{NL}} - (e^{K}\frac{\gamma r^{\alpha^{\rm NL}}}{PA^{\rm NL}}
PA^{NL})^{-1} u^{\alpha^{NL}}}{1 + (\frac{\gamma r^{\alpha^{\rm NL}}}{PA^{\rm NL}}
PA^{NL})^{-1} u^{\alpha^{NL}} - K(\frac{\gamma r^{\alpha^{\rm NL}}}{PA^{\rm NL}}
PA^{NL})^{-1} u^{\alpha^{NL}}} udu \Biggr) \nonumber \\
& &
\label{eq:proof_Lemma4_eq2}
\end{eqnarray}

\noindent We conclude our proof by plugging (\ref{eq:rou1_func}) and (\ref{eq:rou2_func}) into (\ref{eq:proof_Lemma4_eq1}) and (\ref{eq:proof_Lemma4_eq2}). 

\section*{Appendix~D: Proof of Lemma~\ref{lem:laplace_term_UAS1_NLoS_seg2}\label{sec:Appendix-D}}

Considering only NLOS interference, $\mathscr{L}_{I_{r}}\left(\frac{\gamma r^{\alpha^{{\rm {NL}}}}}{PA^{{\rm {NL}}}}\right)$ in the range of $r>d_{1}$ can be derived as

\noindent 
\begin{eqnarray}
\mathscr{L}_{I_{r}}(\frac{\gamma r^{\alpha^{{\rm {NL}}}}}{PA^{{\rm {NL}}}})= \nonumber \\ 
& \hspace{-3cm} {\rm {exp}}\Biggl(-2\pi \lambda \int_{d_{1}}^{\infty} (\frac{1 + (\frac{\gamma r^{\alpha^{\rm NL}}}{PA^{\rm 
NL}} PA^{NL})^{-1} u^{\alpha^{NL}} - K(\frac{\gamma r^{\alpha^{\rm NL}}}{PA^{\rm NL}}
PA^{NL})^{-1} u^{\alpha^{NL}} - (e^{K}\frac{\gamma r^{\alpha^{\rm NL}}}{PA^{\rm NL}}
PA^{NL})^{-1} u^{\alpha^{NL}}}{1 + (\frac{\gamma r^{\alpha^{\rm NL}}}{PA^{\rm NL}}
PA^{NL})^{-1} u^{\alpha^{NL}} - K(\frac{\gamma r^{\alpha^{\rm NL}}}{PA^{\rm NL}}
PA^{NL})^{-1} u^{\alpha^{NL}}}) udu\Biggr) \nonumber \\
& &
\label{eq:proof_Lemma5_eq1}
\end{eqnarray}

\noindent where $(r>d_{1})$. We conclude our proof by plugging (\ref{eq:rou2_func}) into (\ref{eq:proof_Lemma5_eq1}).


\begin{thebibliography}{10}
\bibitem{Report_CISCO}CISCO, ``Cisco visual networking index: Global
mobile data traffic forecast update (2013-2018),'' Feb. 2014. 

\bibitem{LOS_NLOS_Trans}M. Ding, P. Wang, D. L$\acute{\textrm{o}}$pez-P$\acute{\textrm{e}}$rez,
M. Ding, G. Mao and Z. Lin, \textquotedblleft Performance Impact of LoS and NLoS Transmissions in Dense Cellular Networks,\textquotedblright{} \emph{IEEE Trans. on Commun.}, vol. 15, no. 3, pp. 2365-2380, March 2016.

\bibitem{Tutor_smallcell}D. L$\acute{\textrm{o}}$pez-P$\acute{\textrm{e}}$rez,
M. Ding, H. Claussen, and A. H. Jafari, \textquotedblleft Towards 1 Gbps/UE in cellular
systems: understanding ultra-dense small cell deployments,\textquotedblright{} \emph{IEEE
Commun. Surveys and Tutorials}, vol. 17, no. 4, pp. 2078-2101, Fourthquarter 2015.

\bibitem{TR36.872}3GPP, \textquotedblleft TR 36.872, Small cell enhancements
for E-UTRA and E-UTRAN - Physical layer aspects,\textquotedblright{}
Dec. 2013.

\bibitem{TR25.996}3GPP, \textquotedblleft TR 25.996, Spatial channel model for Multiple Input Multiple Output (MIMO) simulations,\textquotedblright{}
Sep. 2012.

\bibitem{Jeff's work 2011}J. G. Andrews, F. Baccelli, and R. K. Ganti,
\textquotedblleft A tractable approach to coverage and rate in cellular
networks,\textquotedblright{} \emph{IEEE Trans. on Commun.}, vol.
59, no. 11, pp. 3122-3134, Nov. 2011.

\bibitem{JSAC_Dhillon}H. S. Dhillon, R. Ganti, F. Baccelli, and J.
G. Andrews, \textquotedblleft Modeling and analysis of K-tier downlink
heterogeneous cellular networks,\textquotedblright{} \emph{IEEE J.
Sel. Areas Commun.}, vol. 30, no. 3, pp. 550-560, Apr. 2012. 

\bibitem{TWC_Singh}S. Singh, H. S. Dhillon, and J. G. Andrews, \textquotedblleft Offloading
in heterogeneous networks: modeling, analysis, and design insights,\textquotedblright{}
\textit{IEEE Trans. on Wireless Commun.}, vol. 12, no. 5, pp. 2484-2497,
May 2013.

\bibitem{book_Haenggi}M. Haenggi, \emph{Stochastic Geometry for Wireless
Networks}. Cambridge University Press, 2012. 

\bibitem{Analysis_random_blockage}T. Bai, R. Vaze, R. W. Heath, \textquotedblleft Analysis
of blockage effects on urban cellular networks,\textquotedblright{}
\textit{IEEE Trans. on Wireless Commun.}, vol. 13, no. 9, pp. 5070-5083,
Sep. 2014.

\bibitem{Wall_pene_indoor}J. Ling, D. Chizhik, \textquotedblleft Capacity
scaling of indoor pico-cellular networks via reuse\textquotedblright ,
\textit{IEEE Commun. Letters}, vol. 16, no. 2, pp. 231-233, Feb. 2012.

\bibitem{related_work_Jeff}X. Zhang, J. G. Andrews, \textquotedblleft Downlink
cellular network analysis with multi-slope path loss models,\textquotedblright{}
\textit{IEEE Trans. on Commun.}, vol. 63, no. 5, pp. 1881-1894, Mar
2015.

\bibitem{Related_work_Health}T. Bai, R. W. Heath Jr., \textquotedblleft Coverage
and rate analysis for millimeter wave cellular networks,\textquotedblright{}\textit{
IEEE Trans. on Wireless Commun.}, vol. 14, no. 2, pp. 1100-1114, Oct.
2014.

\bibitem{related_work_Galiotto}C. Galiotto, N. K. Pratas, N. Marchetti,
L. Doyle, \textquotedblleft A stochastic geometry framework for LOS/NLOS
propagation in dense small cell networks\textquotedblright , {[}Online{]}.
Available: http://arxiv.org/abs/1412.5065

\bibitem{TR36.828}3GPP, \textquotedblleft TR 36.828 (V11.0.0): Further
enhancements to LTE Time Division Duplex (TDD) for Downlink-Uplink
(DL-UL) interference management and traffic adaptation,\textquotedblright{}
Jun. 2012.

\bibitem{SCM_pathloss_model}Spatial Channel Model AHG, \textquotedblleft Subsection
3.5.3, Spatial Channel Model Text Description V6.0,\textquotedblright{}
Apr. 2003. ({[}Online{]}: ftp://www.3gpp.org/tsg\_ran/WG1\_RL1/3GPP\_3GPP2\_SCM/ConfCall-16-20030417)

\bibitem{our_GC_paper_2015_HPPP}M. Ding, D. L$\acute{\textrm{o}}$pez-P$\acute{\textrm{e}}$rez,
G. Mao, P. Wang and Z. Lin, \textquotedblleft Will the area spectral efficiency monotonically
grow as small cells go dense?,\textquotedblright{}\textit{Proc. of IEEE Globecom}, Dec. 2015.

\bibitem{Book_Integrals}I.S. Gradshteyn and I.M. Ryzhik, \emph{Table
of Integrals, Series, and Products (7th Ed.)}, Academic Press, 2007.

\bibitem{Bisection}R. L. Burden and J. D. Faires,\emph{ Numerical
Analysis (3rd Ed.)}, PWS Publishers, 1985.

\bibitem{Rician_Quek}M. Peng and Y. Li and T. Q. S. Quek and C. Wang, \textquotedblleft Device-to-Device Underlaid Cellular Networks under Rician Fading Channels,\textquotedblright{}\textit{
IEEE Trans. on Wireless Commun.}, vol. 13, no. 8, pp. 4247-4259, Aug. 2014.

\bibitem{Rician_Peng}Y. Li and J. Li and J. Jiang and M. Peng, \textquotedblleft Performance analysis of device-to-device underlay communication in Rician fading channels,\textquotedblright{}\textit{Proc. of IEEE Globecom}, Dec. 2013. 

\bibitem{Book_LTE2}H. Holma and A. Toskala, \emph{LTE for UMTS \textendash{}
OFDMA and SC-FDMA Based Radio Access}, John Wiley \& Sons Ltd., 2009.\end{thebibliography}
\end{document}